\documentclass[11pt,draftcls,onecolumn]{IEEEtran}
\IEEEoverridecommandlockouts

\usepackage{amssymb,amsfonts,amsmath,amstext,amsthm,times,cite,enumerate}
\usepackage{pifont}
\usepackage{graphics}
\usepackage{epsfig}
\usepackage{subfig}
\usepackage{wrapfig}
\usepackage{times}
\usepackage{array}

\newcommand{\calB}{{\cal{B}}}

\newcommand{\calS}{{\cal{S}}}
\newcommand{\calU}{{\cal{U}}}

\newcommand{\calX}{{\cal{X}}}

\newcommand{\emb}[1]{\text{\boldmath{$#1$}}}

\newcommand{\Expect}{\textsf{E}}
\newcommand{\indic}[1]{\mbox{$1\hskip -2.5 pt\hbox{I}$}{\{#1\}}}

\newcommand{\QMDP}{$\text{Q}_{\text{MDP}}$}

\newtheorem{remark}{Remark}[section]

\renewcommand{\mathbf}{\boldsymbol}
\title{Sensor Scheduling for Energy-Efficient Target Tracking in Sensor Networks\thanks{This work was funded in part by a grant from the Motorola corporation, a U.S. Army Research Office MURI grant W911NF-06-1-0094 through a subcontract from Brown University at the University of Illinois, a NSF Graduate Research Fellowship, and by a Vodafone Fellowship.}}
\author{George~K.~Atia,~\IEEEmembership{Member,~IEEE}, Venugopal~V.~Veeravalli,~\IEEEmembership{Fellow,~IEEE}, and Jason~A.~Fuemmeler~\IEEEmembership{Member,~IEEE}
\thanks{This work was done at the Coordinated Science Laboratory (CSL), University of Illinois at Urbana Champaign, Urbana IL, and was submitted in part in June 2010 to the Asilomar conference on Signals, Systems, and Computers. Emails: \{atia1,vvv,fuemmele\}@illinois.edu}}
\date{}

\begin{document}
\maketitle

\begin{abstract}
In this paper we study the problem of tracking an object moving randomly through a network of wireless sensors. Our objective is to devise strategies for scheduling the sensors to optimize the tradeoff between tracking performance and energy consumption. We cast the scheduling problem as a Partially Observable Markov Decision Process (POMDP), where the control actions correspond to the set of sensors to activate at each time step. Using a bottom-up approach, we consider different sensing, motion and cost models with increasing levels of difficulty. At the first level, the sensing regions of the different sensors do not overlap and the target is only observed within the sensing range of an active sensor. Then, we consider sensors with overlapping sensing range such that the tracking error, and hence the actions of the different sensors, are tightly coupled. Finally, we consider scenarios wherein the target locations and sensors' observations assume values on continuous spaces.  Exact solutions are generally intractable even for the simplest models due to the dimensionality of the information and action spaces. Hence, we devise approximate solution techniques, and in some cases derive lower bounds on the optimal tradeoff curves. The generated scheduling policies, albeit suboptimal, often provide close-to-optimal energy-tracking tradeoffs.
\end{abstract}

\section{Introduction}
In large networks of inexpensive sensors with small batteries, the sensor nodes are required to operate on limited energy budgets. Sensor management can prolong the lifetime of a sensor network and conserve scarce energy resources. However, inefficient management could result in severe performance degradation.

In this paper, we consider a network of $n$ sensors tracking a single object. The sensors can be turned on or off at consecutive time steps and the goal is to select the subset of sensors to activate at each time step. This problem is challenging due to the inherent tradeoff between the value of information in the sensor measurements and the energy cost, combined with the combinatorial complexity of the decision space.

In previous work \cite{fuemmeler-vvv-tsp-08}, two of the authors considered approximate strategies for \emph{sensor sleeping}, where the sensors are put to sleep to save energy and decisions are made concerning their sleep duration (in time slots). Once in a sleep mode, a sensor would only wake up after its own sleep timer expires. Here, we consider a scheduling variant of the problem which can be thought of as a sleeping problem with an external wake-up mechanism, i.e., sensors can be woken up by external means (e.g. a low-power wake-up radio). At time $k$, the permissible control actions for an $n$-sensor scheduling problem are $n$-dimensional binary vectors, i.e., vectors in  $\{0,1\}^n$ (corresponding to set sensor nodes to activate at each time step), in contrast to vectors in $\mathbb{N}_0^{n_a(k)}$ for the sleeping problem (corresponding to the sleep durations of awake sensors), where $\mathbb{N}_0$ is the set of non-negative integers and $n_a(k)$ the number of awake sensors at time $k$. While this does not address the combinatorial nature of the control space, the simpler structure of the control space for the scheduling problem enables efficient approximate solution methodologies for the more realistic models that we study in this paper.

A significant body of related research work considers sensor management for tasking sensors in dynamically evolving environments. Castanon \cite{castanon} has developed an approximate dynamic programming approach for dynamic scheduling of multi-mode sensor resources for the classification of a large number of unknown objects. The goal is to achieve an accurate classification of each object at the end of a fixed finite horizon by assigning different sensor modes to different objects subject to periodic or total resource usage constraints. Mode allocation strategies are computed based on Lagrangian relaxation for an approximate optimization problem wherein sample-path resource constraints are replaced by expected value constraints. In the context of sensor scheduling for target tracking, information-based approaches  \cite{fisher,hero} have been developed for optimizing tracking performance subject to an explicit constraint on communication costs in a decentralized setting. Williams et al. \cite{fisher} also adopt a Lagrangian relaxation approach to solve a constrained dynamic program over a rolling horizon. There, the combinatorial complexity of the decision space is avoided by first selecting one leader node, followed by greedy sensor subset selection. Other related work on sensor scheduling include  leader-based distributed tracking schemes \cite{zhao,chong}, where at any time instant there is only one sensor active, namely, the leader sensor which changes dynamically as a function of the object state, while the rest of the network is idle.

While previous work focused on developing distributed implementations of efficient sensor scheduling strategies, our goal here is to study the fundamental theory of sensor scheduling for tracking and surveillance applications. Specifically, to explicitly study the fundamental tradeoff between tracking performance and energy expenditure, we define a unified objective function combining tracking and energy costs trading-off the complexity of per-stage costs to better capture the inherent energy-tracking tradeoff. We adopt a bottom-up approach where we consider a range of sensing, motion and cost models with increasing levels of difficulty and devise suboptimal scheduling policies to balance the tradeoff between energy expenditure and tracking performance. In some cases we are also able to derive lower bounds on the optimal energy-tracking tradeoff.

Due to noise and model uncertainties, natural limitations of the measurement devices, or incomplete data about the surroundings, we need to design scheduling policies when the system's state is only partially observable to the controller. Partially-Observable Markov Decision Processes (POMDPs) provide a natural framework for addressing sequential decision problems where the goal is to find a policy (strategy) for selecting actions based on the information available to the controller while addressing both short-term and long-term benefits and costs. Solving POMDPs optimally is generally intractable. For example, the value function for a POMDP with a finite state space depends on information states consisting of conditional probability vectors of dimension equal to the number of states. This has led to a number of POMDP approximations and we refer the reader to Monahan\cite{monahan} and Hauskrecht \cite{Hauskrecht2000} for excellent surveys on approximate methods for stochastic dynamic programming. Usually, no single approximation can be prescribed for all POMDPs, rather approximations can be judiciously used to exploit specific problem structures. In this paper, we use a subset of these approximate solution techniques, including reduced-uncertainty and point-based approximations \cite{Pineau2003,QMDP,spaan2005,porta}. The former assumes that more information would be available to the controller at future time steps, and the latter solves a reduced optimization problem based on a relatively small subset of sampled beliefs about the object's state. We devise different approaches to deal with the aforementioned computational complexity of the decision space. In one approach, instead of solving one large combinatorial problem, we solve a set of simpler subproblems based on the intuition gained from a simplistic sensing model. In another approach, we iteratively sample control actions from a reduced control space based on the sparsity of a reachable belief set combined with point-based value updates.

The remainder of this paper is organized as follows. In Section \ref{sec:problem}, we describe the tracking problem and define the sensing, transition and cost models, as well as the optimization problem, for each of the considered models. In Section \ref{sec:approx_sols} we describe approximate strategies to generate suboptimal scheduling policies. In Section \ref{sec:results}, we  present some experimental results, and finally, in Section \ref{sec:conclusions}, we provide some concluding remarks.

\section{Scheduling Problem}
\label{sec:problem}
In the following we consider different models with increasing level of difficulty. Depending on the structure of the model, we devise approximate methods to address the associated difficulties and generate efficient scheduling policies. For notation, vectors are denoted by bold lower-case letters. Superscript T denotes transposition and the indicator function is written as $\indic{.}$.
%

\subsection{Simple sensing, observation and cost models}
\label{sec:simple_model}
In this model, the network is divided into $n$ distinct cells, one for each sensor. In other words, each cell corresponds to the sensing range of one particular sensor and sensors' ranges do not overlap. A Markov chain with an $(n+1)\times (n+1)$ probability transition matrix $P$ describes the motion of the target through the field of interest. The extra state is for an absorbing termination state of the Markov chain which is reached when the object leaves the network.
%
%
%
%
It is further assumed that all information about the object trajectory is stored at some central unit and is used to determine the scheduling actions for the different sensors.

We let $u_{k,\ell}$ denote the action for sensor $\ell$ at time $k$; $u_{k,\ell} = 1$ if sensor $\ell$ is activated at time $k+1$ and $0$ if the decision is to turn it off. The action vector at time $k$, denoted $\emb{u}_k$, is a binary vector of size $n\times 1$, one decision per sensor. In this simplistic model, we assume that the target is perfectly observable within the cell of an awake sensor or if it reaches the terminal state $\tau$, otherwise it is unobservable. Thus, the observation $s_k$ at time $k$ is defined according to:
\begin{eqnarray}
s_k = \left\{
      \begin{array}{ll}
        b_k, & \hbox{if $b_k\ne\tau$ and $u_{k-1,b_k}=1$;} \\
        \varepsilon, & \hbox{if $b_k\ne\tau$ and $u_{k-1,b_k}=0$;} \\
        \tau, & \hbox{if $b_k=\tau$.}
      \end{array}
    \right.
\label{eq:obs_model_simple}
\end{eqnarray}

\noindent where $\varepsilon$ stands for erasure. The observation model in (\ref{eq:obs_model_simple}) induces a well-defined probabilistic observation model $p(s_k|b_k,\emb{u}_{k-1})$ such that the current observation depends on that actual target location and the scheduling action for the $n$ sensors.


At each time step, the incurred cost is the sum of the energy and the tracking costs. An energy cost of $c\in(0,1]$ per unit time is incurred for every active sensor and a tracking cost of $1$ for each time unit that the object is not observed. Once state $\tau$ is reached the problem terminates and no further cost is incurred. In other words, $\tau$ is an absorbing cost-free state; all $n$ states are transient so that $\tau$ is the only recurrence class of the Markov chain. Hence,

\begin{eqnarray}
g(b_k,\emb{u}_{k-1}) = \indic{b_k\ne\tau}\left(\indic{u_{k-1,b_k}=0}+\sum_{\ell=1}^n c\indic{u_{k-1,\ell}=1}\right)
\label{eq:cost_simple}
\end{eqnarray}

\noindent The parameter $c$ is thus used to tradeoff energy consumption and tracking errors.

\subsection{Overlapping sensors with discrete observations models}
\label{sec:overlap_model}
In this model, we continue to use a discrete model for the target transition but we redefine a new sensing model and cost structure to account for the fact that sensors could have overlapping visibility regions. Within that model we further consider simple and probabilistic sensing.

\subsubsection{\em Overlapping sensors with simple sensing}
In this case, the target is \emph{perfectly} observed within the visibility region of \emph{any} active sensor. Denote by ${\cal R}_{\ell}$ the set of locations in the visibility region of sensor $\ell$ and by ${\cal B}_i$ the set of sensors that observe location $i$. The observation at time $k$ is as follows:
\begin{eqnarray}
s_k = \left\{
      \begin{array}{ll}
        b_k, & \hbox{if $b_k\ne\tau$ and $\exists j\in{\cal B}_{b_k}:u_{k-1,j}=1$;} \\
        \varepsilon, & \hbox{if $b_k\ne\tau$ and $u_{k-1,j}=0,~\forall j\in{\cal B}_{b_k}$;} \\
        \tau, & \hbox{if $b_k=\tau$.}
      \end{array}
    \right.
\label{eq:obsmodel_overlap_simp}
\end{eqnarray}

Therefore, a tracking error is incurred if none of the sensors observing the current target location is active. Redefining the cost structure for this model:
\begin{eqnarray}
g(b_k,\emb{u}_{k-1}) = \indic{b_k\ne\tau}\left(\indic{u_{k-1,j}=0,\forall j\in {\cal B}_{b_k} }+\sum_{\ell=1}^n c\indic{u_{k-1,\ell}=1}\right)
\end{eqnarray}

\subsubsection{\em Overlapping sesnors with probabilistic sensing}
By probabilistic sensing we account for observation uncertainty even if the target is within the visibility region of one or more active sensors. We assume,
\begin{eqnarray}
p(s_k|b_k,\exists j\in{\cal B}_{b_k}:u_{k-1,j}=1)=
      \left\{
        \begin{array}{ll}
          q, & \hbox{$s_k = b_k$;} \\
          \frac{1-q}{|{\cal R}|-1}, & \hbox{$s_k = i,~\forall i\in\cal R$}
        \end{array}
      \right.
\label{eq:obsmodel_overlap_prob}
\end{eqnarray}
\noindent where
\[
{\cal R}=\bigcap_{\substack{j\in{\cal B}_{b_k},\\u_{k-1,j}=1}}R_j\big\backslash\bigcup_{\substack{i\notin{\cal B}_{b_k},\\u_{k-1,i}=1}}R_i.
\]
That is, the observation is uniformly distributed over the remaining locations (other than the true target location) that belong to the visibility regions of the set of awake sensors monitoring the true location $b_k$. If the true target location does not belong to the visibility region of an awake sensor, we naturally exclude the visibility region of that sensor since no measurement is received from such a sensor. When $\cal R$ is a singleton $\{b_k\}$, we set $q=1$. A tracking error is incurred if the target is not directly observed and the uncertainty in the target location cannot be resolved.

\subsection{Continuous observation, continuous state and arbitrary cost models}
\label{sec:cont_model}
In this class of models, the object sensing model allows for an arbitrary distribution for the
observations given the current object location. Tracking cost is modeled as an arbitrary distance measure between the actual and the estimated object location. If we denote the set of possible object locations $\calB$, we have $\calB = m+1$. Note that, in contrast to the simplistic model in \ref{sec:simple_model}, $m$ is different from $n$ since object locations are arbitrary and we no longer assume one location corresponds to the sensing range of one particular sensor. The $(m+1)$-th state again corresponds to a termination state. Furthermore, the target can be moving on a continuous state space in which case $m$ is $\infty$.

If the state space is discrete, then conditioned on the object state $b_k$ at time $k$, $b_{k+1}$ has a probability mass function that is given by the $b_k$-th row of the transition matrix $P$.
%
%
If the state space is continuous, $P$ is a kernel such that $P(x,{\cal Y})$ is the probability that the next object location is in the set ${\cal Y}\subset \calB$ given the current object location is $x$. For simplicity of exposition, we focus on discrete state spaces. 
Also, we omit indexing time whenever the  time evolution is well-understood to avoid cumbersome notation. We consider the following observation model for illustration; however, our approach is fairly general:
\begin{eqnarray}
p(\emb{s}|b,\emb{u}) = \prod_{i=1}^n \left\{\frac{1}{\sqrt{2\pi}} \exp\left(-\frac{1}{2}\left(s_i-\frac{10}{(b-p_i)^2+1}\right)^2\right)\indic{u_i=1}+\delta(s_i-\varepsilon)\indic{u_i=0}\right\}
\label{cont_obs_model}
\end{eqnarray}
where $\emb{s}$ is an $n\times 1$ continuous observation vector with the $i$-th entry, $s_i$, representing the observation of sensor $i$, $p_i, i=1,\ldots,n$, is the position of the $i$-th sensor, $b$ is the target state, and $\varepsilon$ stands for erasure. $\delta(.)$ is the Dirac Delta function. In  (\ref{cont_obs_model}), the observation of an active sensor is Gaussian with a mean received signal strength inversely proportional to the square of the distance between the sensor and the actual target location. The observation of an inactive sensor is just an erasure.

The estimated target location (given the entire history) is denoted by $\hat{b}$. We define the tracking error through an arbitrary bounded distance function $d(b,\hat{b})$ between the actual and the estimated object locations, which can be the Hamming distance $d(b,\hat{b}) = \indic{b\ne\hat{b}}$ or the Euclidean distance for discrete and continuous state spaces, respectively. The control at each time step is the tuple $(\hat{b}_k,\emb{u}_k)$. Since $\hat{b}$ does not affect the state evolution, the optimal value for $\hat{b}_k$ is the value that minimizes the tracking cost over a single time step given history up to time $k$, i.e.,
\begin{equation}
\hat{b}_k = \arg\min_{\hat{b}} E[d(b_k,\hat{b}_k)|I_k]
\end{equation}
\noindent where, $I_k$ denotes the information state, i.e., the total information available to the central controller at time $k$ which is given by
\[
I_k=\{\emb{s}_0,\emb{s}_1\ldots,\emb{s}_k,\emb{u}_0,\emb{u}_1\ldots, \emb{u}_{k-1}\}
\]
\noindent In the case of Hamming cost, it follows that $\hat{b}$ is simply the MAP decision, i.e., $\hat{b} = \arg\max_b p_k(b)$.
\subsection{Optimal scheduling policy}
The design of an \emph{optimal scheduling policy} depends on the history up to time $k$, i.e., the information state $I_k$. However, the posterior probability distribution, $\emb{p}_k = \Pr[b_k|I_k]$, of the target's state given $I_k$ is a sufficient statistic for this class of partially observable processes. The distribution $\emb{p}_k$, also known as belief, summarizes all the information needed for optimal control. The sufficient statistic itself forms a Markov process whose evolution can be obtained through Bayes' rule updates \footnote{Equivalently, for a continuous state space, a sufficient statistic would be $p_k(\calX)=\Pr[b_k\in\calX|I_k]$. The updated belief $p_{k+1}$ can be computed using standard Bayesian non-linear filtering as the posterior measure resulting from prior measure $pP$ and observation $s_{k+1}$.}. For example, the belief update equation for the simplistic model in Section \ref{sec:simple_model} can be written as:
\begin{eqnarray}
\emb{p}_{k+1} = \left\{
      \begin{array}{ll}
        \emb{e}_{\tau}, & \hbox{if $s_{k+1}=\tau$;} \\
        \emb{e}_{b_{k+1}}, & \hbox{if $u_{k,b_{k+1}}=1$;} \\
        \left[\emb{p}_k P\right]_{\{j:u_{k,j}=0\}}, & \hbox{if $u_{k,b_{k+1}}=0$.}
      \end{array}
    \right.
    \label{eq:belief_evolution_simplemodel}
\end{eqnarray}
\noindent where $\emb{e}_i$ is a row vector with a $1$ at the $i$-th entry and $0$ elsewhere. The vector $[\emb{p}_k P]_{\cal S}$ is the probability vector formed by setting the $i$-th entry $[\emb{p}_k P]_i$ of the vector $\emb{p}_k P$ to zero, $\forall i\notin\cal S$, and then normalizing the vector into a probability distribution. The set ${\{j:u_{k,j}=0\}}$ signifies the set of deactivated sensors. In other words, the updated belief for the model in \ref{sec:simple_model}, is a point mass distribution concentrated at $\tau$ if the object exits the network, and concentrated at $b_{k+1}$ if the object is observed. When the object is unobservable, we eliminate the probability mass at all sensors that are awake, since the object cannot be at these locations, and normalize. The multi-valued function in (\ref{eq:belief_evolution_simplemodel}), and equivalent Bayes' updates for the other models, define a transformation $\emb{p}_{k+1}=\phi(\emb{p}_k,s_{k+1},\emb{u}_k)$, mapping the current belief $\emb{p}_k$, the current control vector $\emb{u}_k$, and the future observation $s_{k+1}$, to a future belief.

The policy $\emb{u}_k=\mu_k(I_k)$ is defined as a mapping from information states $I_k$ to control actions $\emb{u}_k$.
The goal is to design a policy that minimizes the expected sum of costs $J$, where,
\begin{equation}
J(I_0,\mu_0,\mu_1,\ldots) = \Expect\left[\sum_{k=1}^{\infty}g(b_k)\middle\vert I_0\right].
\end{equation}
$J$ is well-defined since $g$ is upper bounded by $cn+1$ (regardless of the model) and the expected time till the object exits the network is finite. Note that the termination is inevitable, thus the objective is to reach the termination state with minimal expected cost. Hence, the scheduling policy is the solution of the minimization problem,
\begin{equation}
J^* = \min_{\mu_0,\mu_1,\ldots}J(I_0,\mu_0,\mu_1,\ldots)
\end{equation}

This POMDP problem falls within the class of infinite horizon stochastic shortest path problems. Noting that the termination state is observable, cost-free and absorbing, and that every policy is proper\footnote{A proper policy is a policy that leads to the termination state with probability one regardless of the initial state. In our problem, the scheduling policy does not affect the target motion and all policies are proper in the sense that there is a positive probability that the target will reach the termination state after a finite number of stages.}, a stationary policy $\mu^*(.)$, i.e., one which does not depend on $k$, is optimal in the class of all history-dependent policies and $\emb{p}_k$ is a sufficient statistic for control \cite{bertsekas_book}, i.e., $u^*_k = \mu^*(\emb{p}_k)$, is defined through a time-invariant mapping from the belief space to the action space. $J$ can be written in terms of the sufficient statistic and the optimal policy can be obtained from the solution of the Bellman equation:
\begin{equation}
J(\emb{p})=\min_{\emb{u}\in\{0,1\}^n}E[g(b',\emb{u})|\emb{p},\emb{u}]+\sum_s p(s|\emb{p},\emb{u})J(\phi(\emb{p},s,\emb{u}))
\label{eq:bellman}
\end{equation}
\noindent such that $J(\emb{e}_{\tau})= 0$, where $J(.)$ is the value function for the POMDP, and the expectation is taken over the future state $b'$ which is distributed according to $\emb{p}P$. Note that we removed the time dependence due to the aforementioned time invariance property. For continuous observations, summation over $s$ is replaced by an integration.

\section{Approximate Solutions and Lower bounds}
\label{sec:approx_sols}
There are a number of algorithms for solving POMDPs exactly \cite{sondik,cheng88,kaelbling98}. These algorithms rely on the powerful result of Sondik that the optimal value function for any POMDP can be approximated arbitrarily closely using a set of hyper-planes ($\alpha$-vectors) defined over the belief simplex \cite{sondik}. This fact is the basis for exact value iteration based algorithms, such as the Witness algorithm \cite{cassandra97} for computing the value function.
The result is a value function parameterized by a number of hyper-planes (or vectors) whereby the belief space is partitioned into a finite number of regions. Each vector minimizes the value function over a certain region of the belief space and has a control action associated with it, which is the optimal control for beliefs in its region.

To clarify, in value iteration we generally start with some initial estimate for $J^*$ and repeatedly apply the transformation defined by the right hand side of Bellman equation (\ref{eq:bellman}) until the sequence of cost functions converges. Let $\{\emb{\alpha}_i^{(k)}\}_{i=1}^{|J^{(k)}|}$ denote the set of vectors parameterizing the value function $J^{(k)}$ after $k$ iterations, where $|J^{(k)}|$ is the total number of hyper-planes, and $\emb{\alpha}_i^{(k)}(b)$, which is a hyperplane in the belief space, represents the value of executing the $k$-step policy associated with the $i$-th vector starting from a state $b$. Hence, the value of executing the $i$-th hyperplane policy starting from a belief state $\emb{p}$ is simply the dot product of $\emb{\alpha}_i$ and $\emb{p}$:
\[
J_i^{(k)}(\emb{p}) = \sum_b \emb{p}(b)\emb{\alpha}_i^{(k)}(b) = \emb{p}\cdot\emb{\alpha}_i^{(k)}.
\]

Therefore, the value of the optimal $k$-step policy starting at $\emb{p}$ is simply the minimum dot product over all hyperplanes, i.e.,
\[
J^{*(k)}(\emb{p}) = \min_{\{\emb{\alpha}_i^{(k)}\}} \emb{p}\cdot\emb{\alpha}_i^{(k)}.
\]

Hence, $J^{*(k)}(\emb{p})$ is piecewise linear and concave. Some of the vectors (also known as policy trees) may be dominated by others in the sense that they are not optimal at any region in the belief simplex. Thus, many exact algorithms devise pruning mechanisms whereby a parsimonious representation with a minimal set of non-dominated hyper-planes is maintained \cite{monahan}.

Even though the aforementioned linearity/concavity property makes the policy search a great deal simpler, the exact computation is generally intractable except for relatively small problems. The two major difficulties for exact computation arise from the exponential growth of the vectors with the planning horizon and with the number of observations, and the inefficiencies related to identification of such vectors and subsequently pruning them. Namely, the number of hyper-planes grows double exponentially such that after $k$ steps the number of hyperplanes is $O\left(|{\cal U}|^{|{\cal S}|^k}\right)$, where $|\calU|$ and $|\calS|$ denote the cardinality of the control and observation spaces, respectively. Equivalently, the number of hyperplanes per iteration grows as:
\[
|J^{(k+1)}|= O\left(|{\cal U}||J^{(k)}|^{|{\cal S}|}\right).
\]
This has led to a number of approximations and suboptimal solutions techniques trading off solution quality for speed.

\begin{remark}
\label{rem:intractability_issues}
The intractability of the optimal solution for our problem is primarily due to the following reasons:

\begin{enumerate}[(i)]
 \item\label{item_belief} The cost function is minimized over the simplex of probability distributions, i.e., the $(m-1)$-dimensional belief simplex for $m$-state discrete state-space models, and the space of probability density functions for continuous state-space models.
 \item\label{item_action} The exponential explosion of the action space with the number of sensors ($2^n$ actions).
 \item\label{item_obs} The exponential growth of the $\alpha$-vectors with the planning horizon and with the number of observations, especially for continuous observation models.
\end{enumerate}
\end{remark}

\subsection{Approximate solutions}
\label{subsec:approx_sols}
In this section, we outline our approximate solution methodologies for the different models introduced in Section \ref{sec:problem}. First, we consider approximations where it is assumed that more information becomes available to the controller at future time steps. Policies based on the assumption that uncertainty in the current belief state will be gone after the next action were first introduced within the artificial intelligence community and known as \QMDP{} policies ~\cite{cassandra97,QMDP}. We show that under an observable-after-control assumption, our sensor scheduling problem decomposes into $n$ simpler subproblems, one subproblem per sensor, \emph{for the simplistic model} of \ref{sec:simple_model}. These subproblems can then be solved exactly using policy iteration \cite{bertsekas_book}. Furthermore, in this case, the \QMDP{} solution gives us a lower bound on the optimal tracking-energy tradeoff. Unfortunately, this natural decomposition does not extend to the other class of models due to the inherent coupling of their tracking errors. However, based on intuition gained from the simplistic model, we artificially decouple the scheduling problem for those models and individually learn the tracking costs corresponding to each subproblem under the aforementioned \QMDP{} assumption. This approach combines \QMDP{} with reinforcement learning \cite{kaelbling96}.

\begin{figure*}
\centering
\includegraphics[width=10cm]{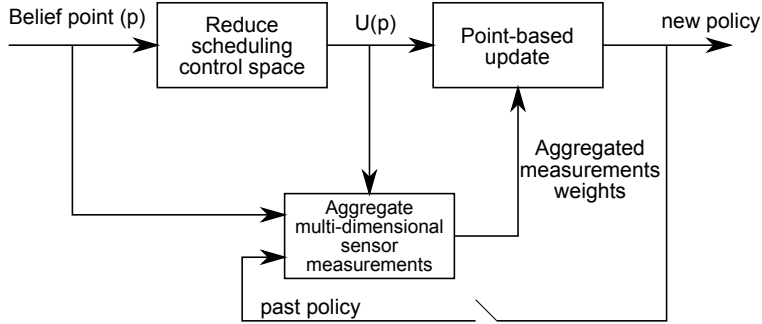}
\caption{Structure of the point-based scheduling approximation}
\label{fig:algorithm}
\end{figure*}

Second, we develop sensor scheduling strategies based on point-based approximations. Despite the fact that the generated \QMDP{} based policies perform reasonably well, generally the resulting policies would not take actions to gain information (an effect of the observable-after-control assumption), leading to situations wherein the belief state does not get updated appropriately. Furthermore, while decoupling the scheduling problem provides close-to optimal performance for uncoupled or lightly-coupled sensing and tracking models (see Section \ref{sec:results}), it might come at the expense of reduction in solution quality for more realistic or heavily-coupled models. To that end, we develop point-based approximate scheduling policies. While our previous approach reduced complexity via decoupling and learning, the key idea here is to optimize the value function only for a small set of reachable beliefs $\cal P$ and not over the entire belief simplex. Point-based methods have shown great potential for solving large scale POMDPs mostly for robotic applications \cite {Hauskrecht2000,Roy2003, Pineau2003, spaan2005}. Pineau et al. \cite{Pineau2003} proposed point-based value iteration (PBVI) which performs point-based backups only at a discrete set of reachable belief points, that can be actually encountered by interacting with the environment. Developing a class of point-based algorithms, which mostly differ in the way the subset of belief points is chosen and the execution order of the backup operations over the selected belief points, has been the focus of recent algorithm-development research targeting large scale POMDPs. Perseus \cite{spaan2005} is one such randomized point-based algorithm that maintains a fixed set of belief points. There, backup speedups can be obtained by exploiting the key observation that a single backup may improve the value of many belief points simultaneously. These algorithms were designed to deal with large state spaces, yet, two extra difficulties in the scheduling problem arise from the size of the action space $2^n$ (for all models) and the observation space (for the models in Sections \ref{sec:cont_model}). Regarding the dimensionality of the action space, we devise a strategy to sample actions based on the support of the beliefs and the sparse structure of the transition models. Intuitively speaking, an object can only move from one side of the network to the other side within time constraints rendering exponentially many scheduling actions irrational at certain times. Hence, instead of performing full updates including $2^n$ actions, we perform the minimization over a reduced control space ${\cal U}(\emb{p})$ for every $\emb{p}\in{\cal P}$ (see Section \ref{sec:sample_actions}). When dealing with continuous or large observations, we combine that with a methodology that aggregates observations and uses aggregate observations for value iteration updates (Section \ref{sec:obs_aggregation}).
At the core of the algorithm we use Perseus \cite{spaan2005}, a variant of PBVI \cite{Pineau2003}, whereby value iteration updates are not carried out for every sampled belief. Instead, the values for many belief points are improved simultaneously in one update. Fig. \ref{fig:algorithm} depicts the structure of our point-based approximation, combining control space reduction and observation aggregation with point-based updates.

\subsection{\QMDP{} based scheduling policies}
\label{sec:qmdp}
Next, we consider our first class of policies based on the \QMDP{} reduced future uncertainty assumption. 
First, we consider the simplistic model in Section \ref{sec:simple_model}, then we use the intuition we developed from this model to devise similar policies for the other models. Since the POMDP is a stochastic shortest path problem with an absorbing cost-free termination state, and the expected termination time is finite, the cost-to-go function for a given belief can be written as the minimum of the dot product of the belief vector and a set of hyper-planes ($\alpha$ vectors):
\begin{align}
J(\emb{p}) &= \min_{\{\emb{\alpha}_i\}} \sum_{b} \alpha_i(b)\emb{p}(b)\nonumber\\
&= \min_{\emb{u}\in\{0,1\}^n}\Bigg\{\sum_{i=1}^n [\emb{p}P]_i\left(\indic{u_i=0}+\sum_{\ell=1}^n c \indic{u_{\ell}=1}\right) \nonumber\\
&\qquad\qquad\qquad + \sum_{s\in\{1\ldots n,\varepsilon\}}\min_{\emb{\alpha}_i}\sum_{b'}p(s|u,b')\sum_b p(b'|b)\emb{p}(b)\emb{\alpha}_i(b')\Bigg\}
\label{eq:bellman_simple_model}
\end{align}
\noindent where $\{\emb{\alpha}_i\}$ is the set of hyperplanes constituting the value function $J$. In essence, the complexity of the Bellman equation (\ref{eq:bellman_simple_model}) stems from the evolution of the belief $\emb{p}_k$ in (\ref{eq:belief_evolution_simplemodel}). We can see why (\ref{eq:bellman_simple_model}) is hard to analyze if we further divide the second term in the summation into two terms depending on whether there is observability or there is an erasure,
\begin{align}
J(\emb{p})&= \min_{\emb{u}\in\{0,1\}^n}\Bigg\{\sum_{i=1}^n [\emb{p}P]_i\left(\indic{u_i=0}+\sum_{\ell=1}^n c \indic{u_{\ell}=1}\right) \nonumber\\
&\qquad\qquad + \sum_{b'}\indic{u_{b'}=1}[\emb{p}P]_{b'}\min_{\{\emb{\alpha}_i\}}\emb{\alpha}_i(b')
+\min_{\{\emb{\alpha}_i\}}\sum_{b'}\indic{u_{b'}=0}[\emb{p}P]_{b'}\emb{\alpha}_i(b')\Bigg\}
\label{eq:intuition_alphas}.
\end{align}

To further clarify we observe that:
\begin{eqnarray}
\sum_s p(s|\textbf{u},p) J(\emb{p}_1) &=& \sum_{i=1}^n \indic{u_i=1}[\emb{p}P]_i J(\emb{e}_i)
+ \sum_{i=1}^n \indic{u_i=0}[\emb{p}P]_i J([\emb{p}P]_{\{j:u_j=0\}})
\label{eq:intuition}
\end{eqnarray}
\noindent and the minimization problem is coupled across the sensors as the second term in (\ref{eq:intuition}), which is due to non-observability, depends on the action vector $\emb{u}$. The action of one sensor affects belief evolution therefore coupling the problem across sensors. Now, if we make the assumption that perfect observations would be available to the controller after taking a scheduling action, we obtain an approximate surrogate function which can be used to generate a suboptimal scheduling policy. Namely, we replace $p(s|u,b')=\delta(s-b')$ in (\ref{eq:bellman_simple_model}). We get
\begin{align}
J(\emb{p}) &= \min_{\emb{u}\in\{0,1\}^n}\Bigg\{\sum_{i=1}^n [\emb{p}P]_i\left(\indic{u_i=0}+\sum_{\ell=1}^n c \indic{u_{\ell}=1}\right)+\sum_{b'} [\emb{p}P]_{b'}\min \emb{\alpha}_i\cdot \emb{e}_{b'}\Bigg\}\nonumber\\
&= \min_{\emb{u}\in\{0,1\}^n}\Bigg\{\sum_{i=1}^n [\emb{p}P]_i\left(\indic{u_i=0}+\sum_{\ell=1}^n c \indic{u_{\ell}=1}\right)+\sum_{b'} [\emb{p}P]_{b'}J(\emb{e}_{b'})\Bigg\}.
\label{eq:surrogate_QMDP}
\end{align}

The terms in the summation in (\ref{eq:surrogate_QMDP}) only depend on the control action for each sensor. Furthermore, the belief evolution is independent of the scheduling action, wherefore the approximate recursion in (\ref{eq:surrogate_QMDP}) decomposes into separable terms, one per sensor. Hence, the value function and the scheduling policy for sensor $\ell$, under the observable-after-control assumption, can be obtained from the solution of per-sensor Bellman equation:
\begin{align}
J^{(\ell)}(\emb{p}) &= \min_{u_{\ell}\in\{0,1\}}\Bigg\{\sum_{i=1}^n [\emb{p}P]_i\left(\indic{u_i=0}+\sum_{\ell=1}^n c \indic{u_{\ell}=1}\right)+\sum_{b'} [\emb{p}P]_{b'}J^{(\ell)}(\emb{e}_{b'})\Bigg\}.
\end{align}

The POMDP problem is now decomposed into $n$ separate simpler subproblems such that the total cost function is the sum of the per-sensor cost function while the overall scheduling policy is the per-sensor policies applied in parallel. Each subproblem can be easily solved using standard policy iteration \cite{bertsekas_book} with a simple minimization over a binary control action.

Fundamentally, for the simplistic model, we were able to decompose the problem into $n$ simpler subproblems due to the separability of the tracking cost into per-sensor costs. Note that the problem is still coupled due to the belief evolution in (\ref{eq:belief_evolution_simplemodel}) yet that coupling is resolved under the observable-after-control assumption.

While separability holds for the simplistic model, this is not the case for the other models. Hence, we devise a strategy where we artificially decouple the problem into $n$ simpler subproblems. To this end, we perform Monte Carlo simulations to determine appropriate values for the per-sensor tracking cost corresponding to each subproblem. For example, consider the continuous observation model of Section \ref{sec:cont_model}. For simplicity of exposition, assume a discrete state space model with $m$ possible object locations. In this case, we define a surrogate value function for the $\ell$-th subproblem as follows:
\begin{equation}
J^{\ell}(\emb{p}) = \min_{u}\Bigg\{\indic{u=0}\sum_{i=1}^m \emb{p}(i) T(i,\ell) + \indic{u=1}\sum_{i=1}^m c[\emb{p}P]_i + \sum_{i=1}^m [\emb{p}P]_i J^{\ell}(\emb{e_i})\Bigg\}~~~~\ell = 1,\ldots,n
\label{eq:artif_decomp}
\end{equation}
\noindent where $T(i,\ell)$ captures the contribution of the $\ell$-th sensor to the total tracking error when the target's previous state is $i$ and is obtained via Monte Carlo simulations. Namely, the expected tracking cost can be evaluated by repeatedly simulating our system from time $k-1$ to time $k$ while changing the state of the $\ell$-th sensor. Similarly,  (\ref{eq:artif_decomp}) can be generalized for continuous state spaces.

Even though the \QMDP{} assumption leads to a separable problem and provides a lower bound on the optimal energy-tracking tradeoff for the simplistic model as we elaborate in Section \ref{sec:lower_bounds}, the resulting scheduling policies are myopic, unlike the sleeping policies in \cite{fuemmeler-vvv-tsp-08}. This follows from the fact that under an observable-after-control assumption, the future cost term is independent of the control vector $\emb{u}$. Therefore, we consider more efficient, albeit more difficult, point-based approximations in the next section.

 \subsection{Point-based approximate policies}
In the previous section, we described \QMDP{} based policies, whereby issues (\ref{item_belief}) and (\ref{item_obs}) in Remark \ref{rem:intractability_issues} are resolved since we only needed to solve the underlying Markov Decision Process to describe the full approximate surrogate function. Decoupling the problem into one-per-sensor subproblems (naturally or artificially) further enabled us to address issue (\ref{item_action}). Yet, we just argued in Sections \ref{subsec:approx_sols} and \ref{sec:qmdp} that the resulting scheduling policies are myopic and generally do not take control actions to gain information.

To that end, we develop point-based approximate scheduling policies. Instead of reducing complexity via artificial decoupling and learning, the key idea here is to optimize the value function only for specific reachable sampled beliefs and not over the entire belief simplex (addressing issue (\ref{item_belief}) in Remark \ref{rem:intractability_issues}). Such techniques have shown great potential for solving large scale POMDPs while significantly reducing complexity. Due to the large size of the control space, we also devise strategies to sample actions exploiting the sparsity of the beliefs and the problem structure (to address issue (\ref{item_action})). Moreover, observation aggregation is used for continuous observation models. Furthermore, since Perseus updates are not carried out for every sampled belief and multiple belief points are improved simultaneously, the number of $\alpha$ vectors grows modestly with the number of iterations. This addresses issue (\ref{item_obs}) in Remark \ref{rem:intractability_issues}.

For completeness we first briefly outline the steps of Perseus and refer the reader to \cite{spaan2005,porta} for further details. Later, we discuss specific variations to the algorithm to address the dimensionality of the action and the observation spaces.

\noindent\textbf{One iteration of Perseus}
\begin{enumerate}
\item\label{sample_step} Sample a set of belief points $\cal P$. We obtain these beliefs by simulating the target motion through the field taking random actions and generating observation according to the observation models in (\ref{eq:obs_model_simple}), (\ref{eq:obsmodel_overlap_simp}), (\ref{eq:obsmodel_overlap_prob}), and (\ref{cont_obs_model})
\item Sample a belief point $\emb{p}\in{\cal P}$ at random and compute the backup using (\ref{eq:backup_a}) and (\ref{eq:backup_b}),
\begin{subequations}
\begin{equation}
\emb{\alpha} = \arg\min_{\{\emb{\alpha}_{\emb{u}}^{\emb{p}}\}_{\emb{u}\in\calU}}\emb{p}\cdot \emb{\alpha}_{\emb{u}}^{\emb{p}}
\label{eq:backup_a}
\end{equation}
\noindent where
\begin{equation}
\emb{\alpha}_{\emb{u}}^{\emb{p}} = g(b,\emb{u})+\sum_s p(s|\emb{u},\emb{p})\min_{\emb{\alpha}_i^{(k)}}\phi(\emb{p},\emb{u},s)\cdot\emb{\alpha}_i^{(k)}
\label{eq:backup_b}
\end{equation}
\label{eq:backup}
\end{subequations}

\item If $\sum_b \emb{p}(b)\emb{\alpha}(b)\leq J^{(k)}(\emb{p})$ then add new $\emb{\alpha}$ to $J^{(k+1)}$ otherwise keep old hyperplane
\item If $\{\emb{p}\in{\cal P}:J_{k+1}(\emb{p}) > J^{(k)}(\emb{p}) \}=\emptyset$, i.e., the empty set, iteration is complete otherwise repeat from step \ref{sample_step}
\end{enumerate}

\begin{figure*}
  \centering
  \subfloat[]{\label{fig:perseus3}\includegraphics[width=0.3\textwidth]{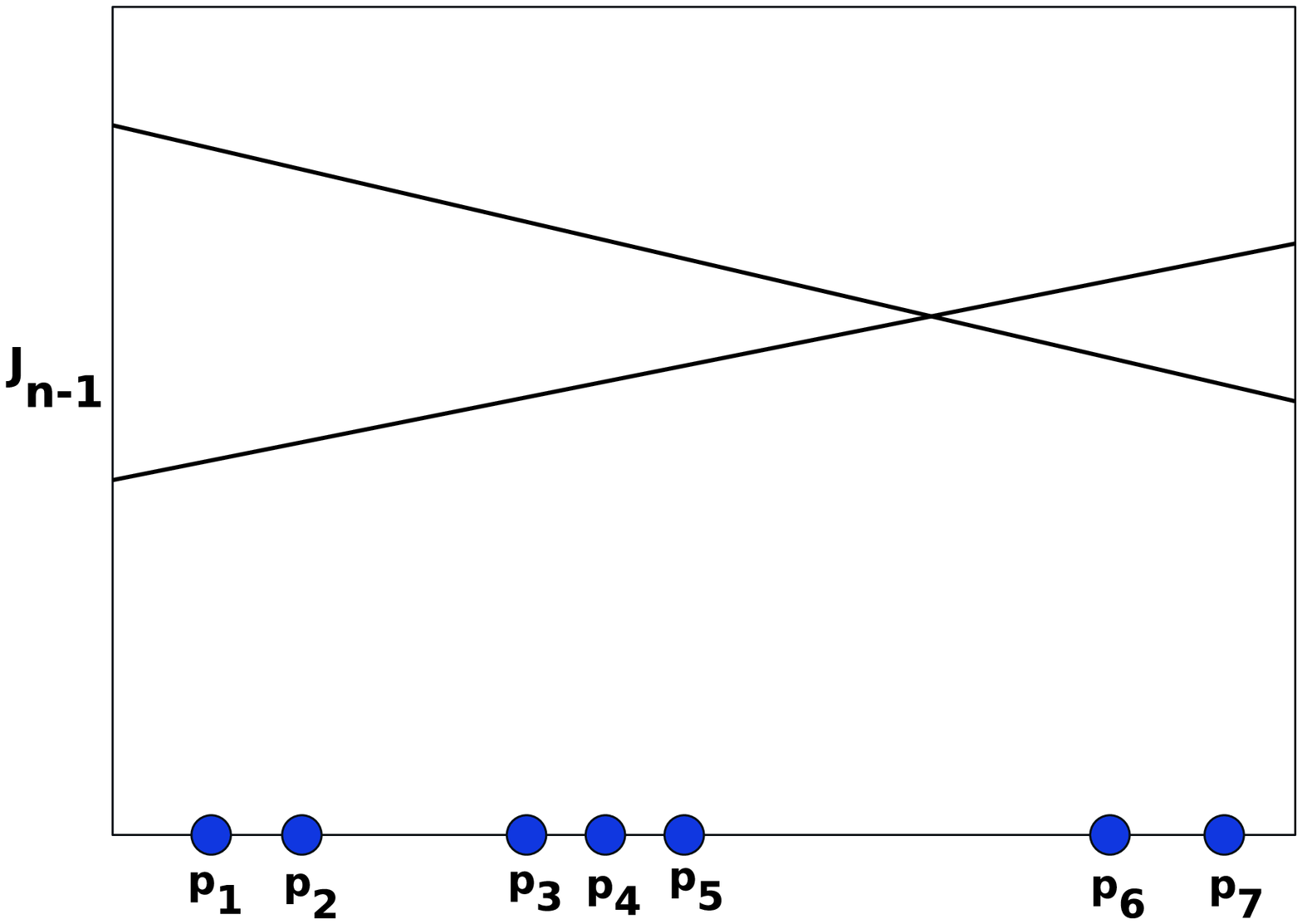}} \hspace{0.25in}
  \subfloat[]{\label{fig:perseus4}\includegraphics[width=0.3\textwidth]{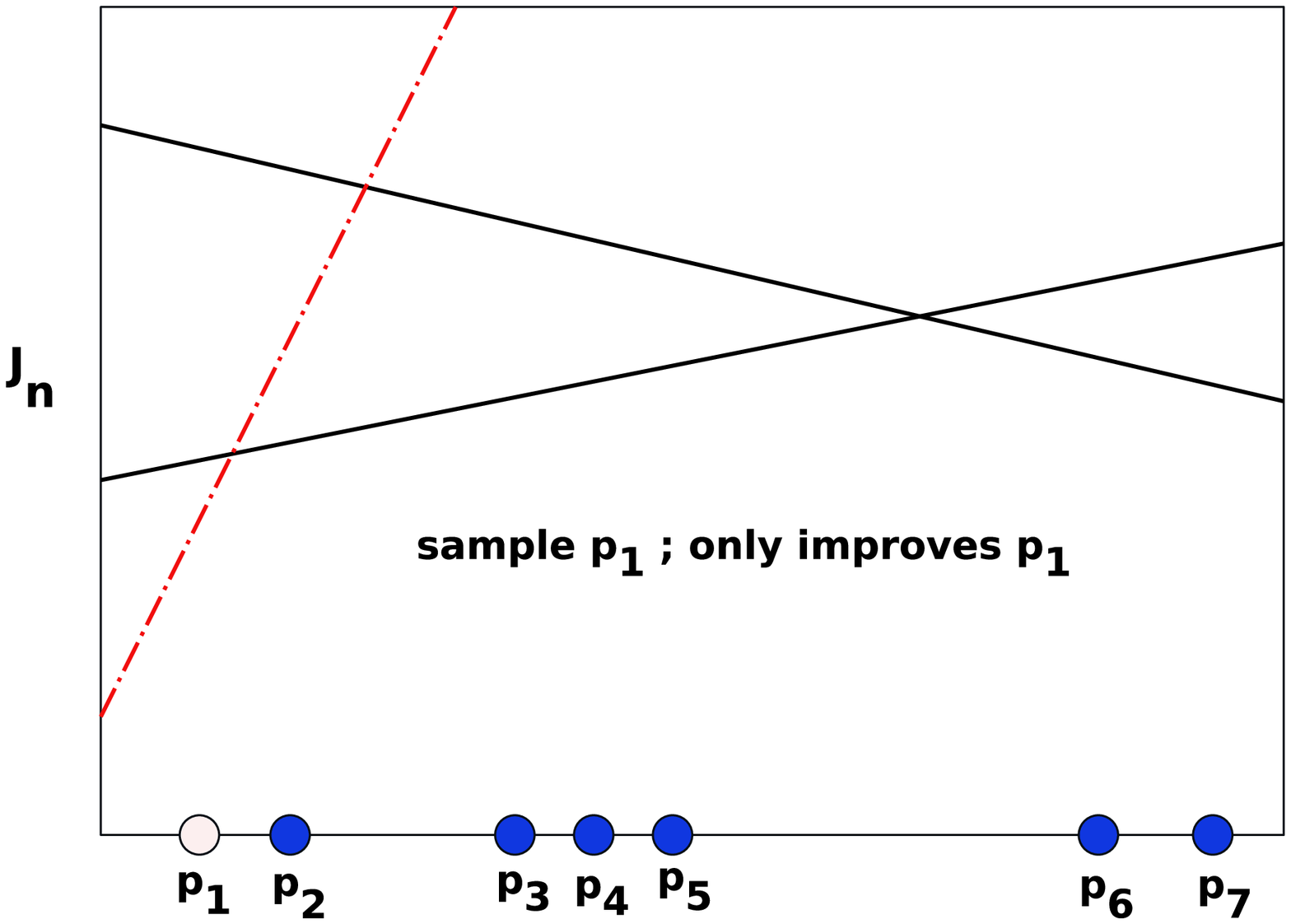}}\hspace{0.25in}\\
  \subfloat[]{\label{fig:perseus3}\includegraphics[width=0.3\textwidth]{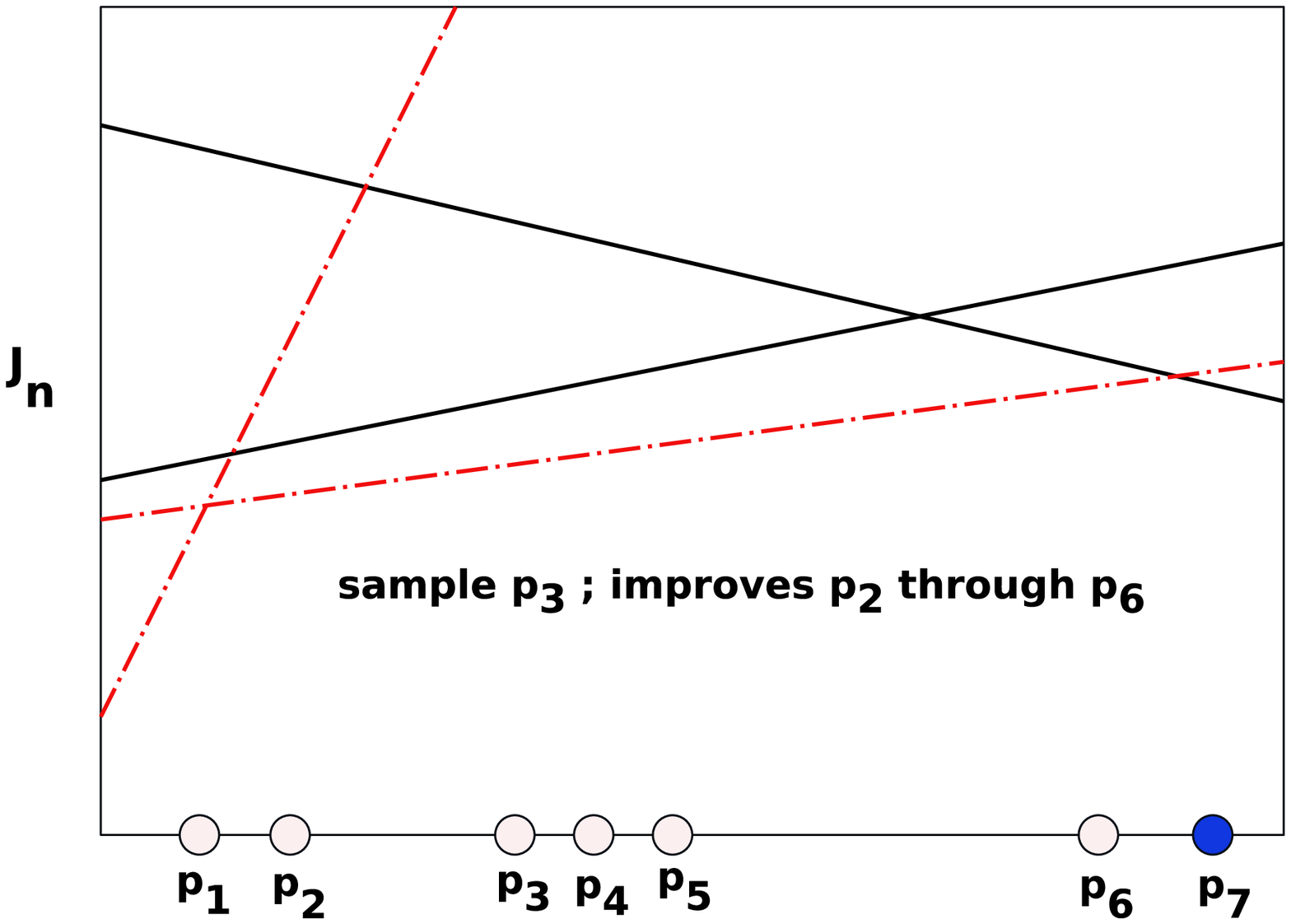}} \hspace{0.25in}
  \subfloat[]{\label{fig:perseus4}\includegraphics[width=0.3\textwidth]{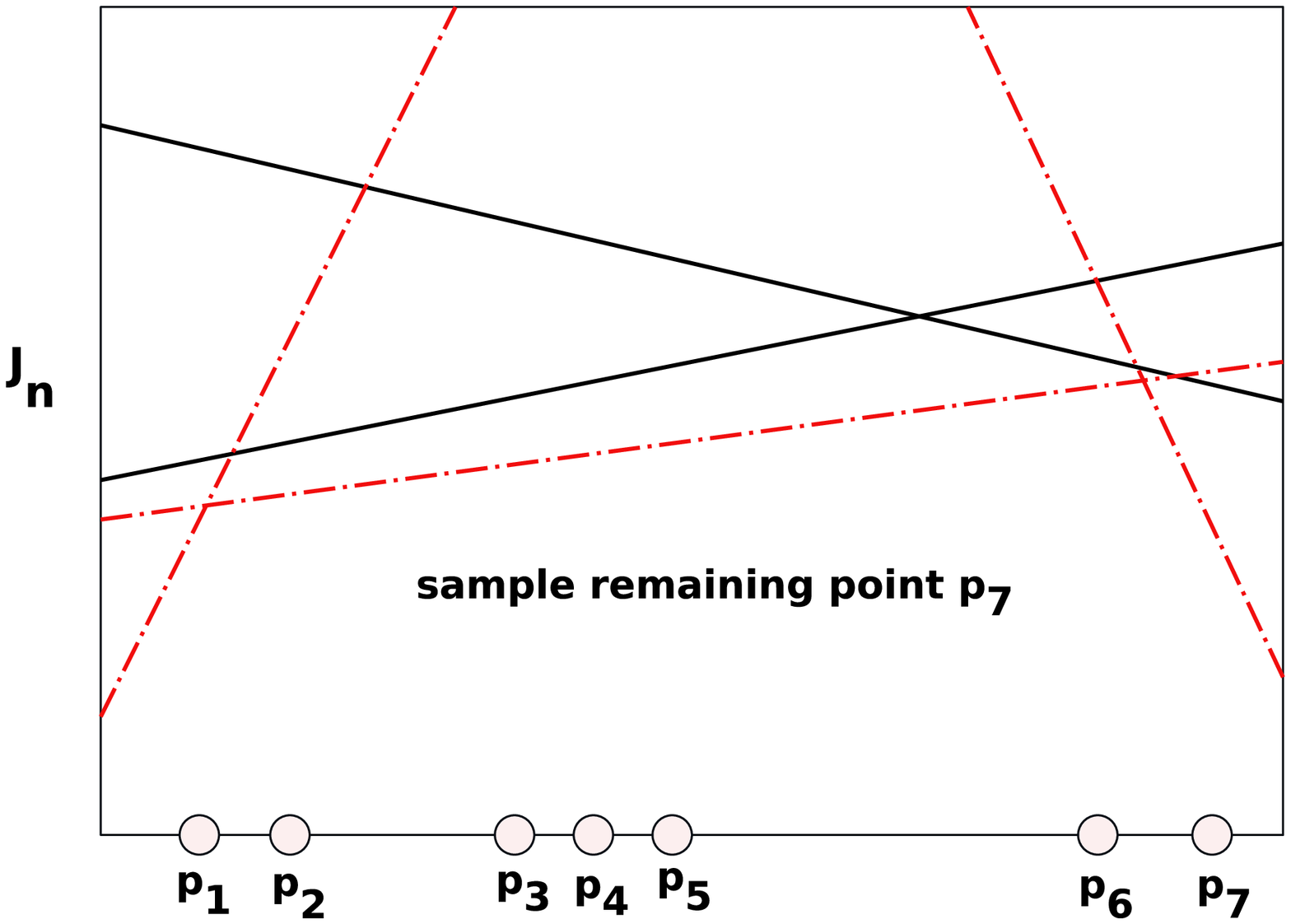}}\hspace{0.25in}
  \subfloat[]{\label{fig:perseus5}\includegraphics[width=0.3\textwidth]{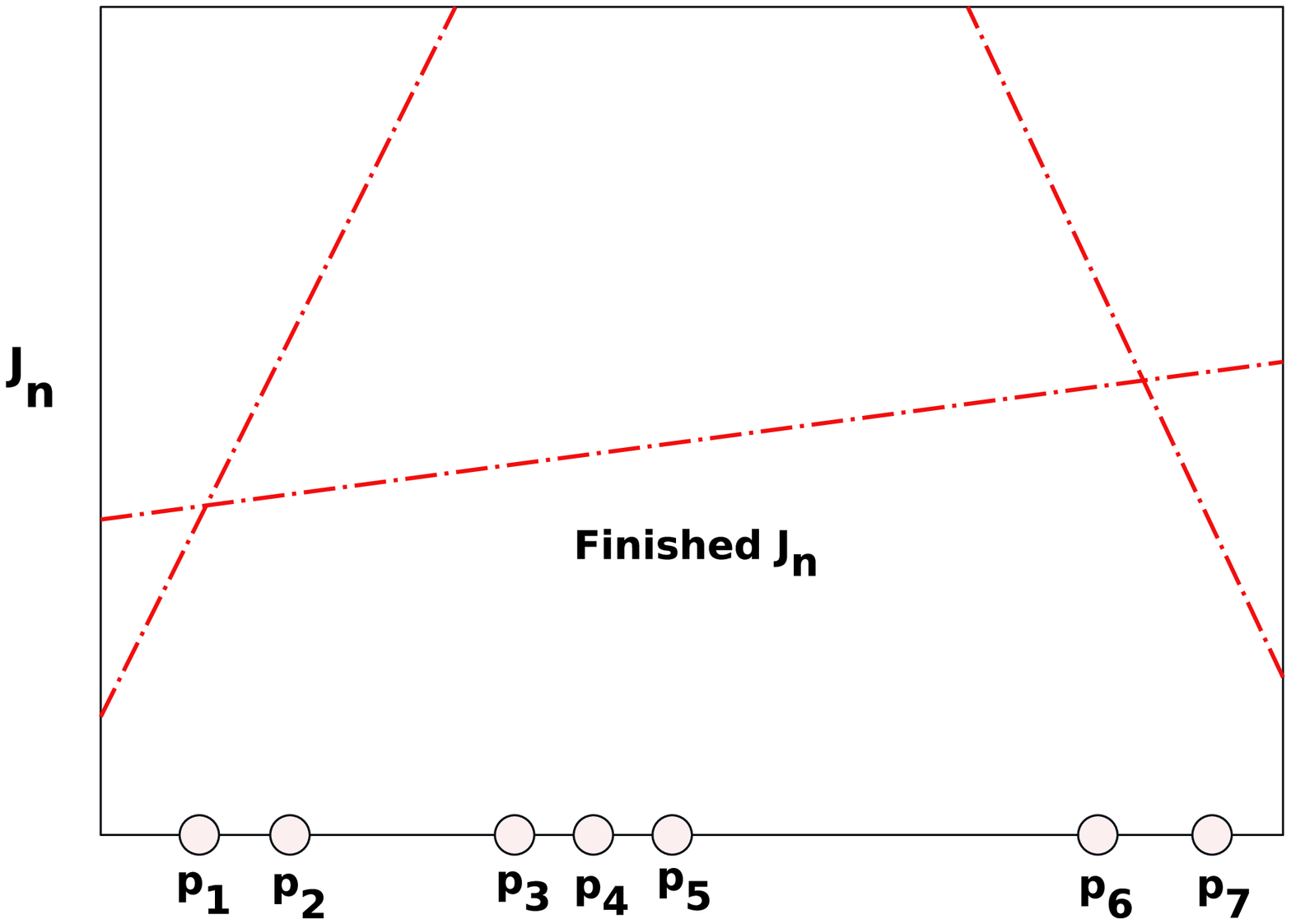}}
  \caption{One iteration of Perseus illustrating the progress of the algorithm. The x-axis represents the belief space with circles representing the sampled belief set ${\cal P}=\{p_1,\ldots,p_7\}$. The y-axis is the value function at consecutive iterations, i.e. $J^{k-1}$ and $J^{k}$. Solid lines represent the hyper-planes in the $(k-1)$-th iteration and dashed lines represent the newly added hyper-planes during the $k$-th iteration. (a) The initial value function $J^{k-1}$; (b) $p_1$ is randomly selected and a new $\alpha$ vector is added to $J^{k}$. This update step only happens to improve $p_1$. Dark circles represent belief points which did not yet improve; (c) $p_3$ is sampled and a new hyperplane is added which improves the value for $p_2$ through $p_6$; (d) Only $p_7$ did not improve, thus $p_7$ is sampled and a new hyperplane is added to $J^{(k)}$; (e) All belief points improved, $J^{(k)}$ is computed, the iteration ends.}
  \label{fig:perseus_illustration}
\end{figure*}

Fig.\ref{fig:perseus_illustration} illustrates the progress of one iteration of Perseus. The x-axis represents the belief space with circles representing the sampled belief set ${\cal P}=\{p_1,\ldots,p_7\}$. The y-axis is the value function at consecutive iterations, i.e. $J^{k-1}$ (solid lines) and $J^{k}$ (dashed lines). The figure displays the $\alpha$ vectors and different steps illustrating the progress of the algorithm. The algorithm selects a belief point at random and updates the value function for that belief. Then a new update is carried out for a belief point randomly selected from the set of remaining beliefs, i.e., beliefs which did not improve in the previous step. The algorithm repeats till all belief points are updated. Solid lines represent the hyper-planes in the $(k-1)$-th iteration and dashed lines represent the newly added hyper-planes during the $k$-th iteration.
\noindent In a way, the Perseus updates in POMDPs are the counterpart of asynchronous dynamic programming for MDPs \cite{bertsekas_book} since the order of backup of the belief points is arbitrary and does not require full sweeps over the entire sampled belief set.
\bigbreak
\subsubsection{\textbf{Sampling actions based on the support of the belief}}
\label{sec:sample_actions}
Note that the update equation (\ref{eq:backup}) involves a minimization over all control actions in $|\cal U|$. Even though one iteration of the algorithm is linear in the cardinality $|{\cal U}|$  of the control space, $|{\cal U}|$ itself is exponential in the number of sensors rendering the minimization infeasible for a relatively large sensor network.

The idea here is to exploit the structure of the scheduling/tracking problem. Since the target transition model is naturally sparse, we predict relatively small uncertainty regions for the target state at future time steps. More specifically, for every belief point in $\cal P$, we use prior information about the target transition model to project the future state of the target. This is particularly useful when the current belief vector is sparse leading to more restricted uncertainty regions. Subsequently, we restrict our attention to a \emph{significant} subset of sensors, that is, sensors of relevance to the particulars of the uncertainty region. Hence, we only consider scheduling actions involving scheduling different combinations of a reduced number of sensors which considerably reduces the control space for every belief in $\cal P$. If the number of significant sensors is still large, we randomly sample actions from the reduced control space. Note that the same intuition extends to more complex motion models wherein information about target speed, maneuver, and acceleration can be factored in to define the future uncertainty regions. Hence, instead of performing full updates including $2^n$ actions, we perform the minimization over a reduced control space for every $\emb{p}\in{\cal P}$. Specifically, we redefine the point update equation as:

\begin{equation}
\emb{\alpha} = \arg\min_{\{\emb{\alpha}_{\emb{u}}^{\emb{p}}\}_{\emb{u}\in\calU(\emb{p})}}\emb{p}\cdot \emb{\alpha}_{\emb{u}}^{\emb{p}}
\label{eq:backup_modified}
\end{equation}

\noindent where $\calU(\emb{p})$ designates the reduced control space for the belief vector $\emb{p}$.

Note that, future iterations of the algorithm involving a particular belief point, ensure sufficient sampling to relevant control actions in the reduced control space. This approach is well suited to Perseus wherein the value for every belief point is guaranteed to improve over consecutive stages of the algorithm. It is worth mentioning that the observation and the cost models need to be computed on the fly for each sampled control action during the algorithm implementation.
\bigbreak
\subsubsection{\textbf{Observation aggregation}}
\label{sec:obs_aggregation}
The point update equation (\ref{eq:backup}) involves back-projecting all hyper-planes in the current iteration one step from the future and returning the vector that minimizes the value of the belief. Since this involves computing a cross sum by enumerating all possible combinations of alpha vectors for the different observations, a number of vectors which is exponential in the number of the observations is generated at each stage. The recursion has to be redefined to address continuous observation models. Looking carefully at (\ref{eq:backup}), it is not hard to see that if different observations map to the same minimizing hyperplane, then they can be aggregated \cite{poupart}. Hence, if we can partition the observation space into regions that map to the same hyperplane (possibly non contiguous), the continuous model is reduced to a corresponding discrete model. Integration is replaced by a summation over these partitions and the weighing probabilities are obtained by integrating the conditional density over these partitions. This is clarified in the following:
\begin{align}
\int_s\min_{\emb{\alpha}_i}\sum_{b'}p(s|u,b')\sum_b p(b'|b)\emb{p}(b)\emb{\alpha}_i(b')~ds &= \sum_j\int_{{\cal S}_j}\sum_{b'}p(s|u,b')\sum_b p(b'|b)\emb{p}(b)\emb{\alpha}_j(b')~ds\nonumber\\
&=\sum_j\sum_{b'}[\emb{p}P]_{b'}\emb{\alpha}_j(b')\int_{{\cal S}_j}p(s|\emb{u},b')~ds\nonumber\\
&=\sum_j\sum_{b'}[\emb{p}P]_{b'}\Pr[{\cal S}_j|\emb{u},b']\emb{\alpha}_j(b').
\end{align}

To find the regions of aggregate observations, we need to solve for the boundaries, i.e., for each pair $(i,j)$ of $\alpha$ vectors we need to solve for $\emb{s}$:
\begin{eqnarray}
\emb{\alpha}_i \cdot \phi(\emb{p},\emb{u},\emb{s})=\emb{\alpha}_j \cdot \phi(\emb{p},\emb{u},\emb{s})
\end{eqnarray}
\noindent where $\phi(\emb{p},\emb{u},\emb{s})= \emb{p}_1(b') \propto \sum_b \emb{p}(b)p(\emb{s}|b',\emb{u})p(b'|b)$

Hence, we need to solve:
\begin{equation}
\sum_{b'}(\emb{\alpha}_i(b')-\emb{\alpha}_j(b'))[\emb{p}P]_{b'}\exp\left\{-\frac{1}{2}\sum_{i:u_i=1}(s_i-\frac{10}{(b'-p_i)^2+1})^2\right\}=0
\label{eq:boundaries}
\end{equation}

After solving for the boundaries, we can readily define the regions:
\begin{equation}
{\cal S}_{j^*} = \{\emb{s}|j^* = \arg\max_j \emb{\alpha}_j\cdot\phi(\emb{p},\emb{u},\emb{s})\}
\end{equation}

Now the update step is simply:
\begin{equation}
J(\emb{p}) = g(\emb{p},\emb{u}^*)+ \sum_j \sum_{b'}[\emb{p}P]_{b'}\Pr[{\cal S}_j|\emb{u}^*,b']\emb{\alpha}_j(b')
\end{equation}
\noindent where
\[
\Pr[{\cal S}_j|\emb{u}^*,b'] = \int_{\emb{s}\in{\cal S}_j} p(\emb{s}|\emb{u}^*,b')d\emb{s}.
\]
Finding a closed form analytical solution for (\ref{eq:boundaries}) is not feasible. Instead, we use Monte-Carlo simulations to solve for the boundaries and get estimates of the weighing probabilities by sampling observations from $p(s|u,b')$ for different combinations of actions and target states.

\subsection{Lower bounds}
\label{sec:lower_bounds}
We are able to derive lower bounds on the energy-tracking tradeoff for the simple as well as the continuous Gaussian observation models. \emph{For the simple model}, the \QMDP{} value function is itself a lower bound on the expected total cost since more information is available to the controller at future time steps given the reduced uncertainty assumption. To further clarify, observe that if we interchange the order of minimization and summation in the last term of (\ref{eq:intuition_alphas}), we obtain a lower bound on the optimal cost to go function. Hence, a lower bound can be obtained from the solution of the following equation:
\begin{align}
J(\emb{p})&= \min_{\emb{u}\in\{0,1\}^n}\Bigg\{\sum_{i=1}^n [\emb{p}P]_i\left(\indic{u_i=0}+\sum_{\ell=1}^n c \indic{u_{\ell}=1}\right) \nonumber\\
&\qquad\qquad\qquad+ \sum_{b'}\indic{u_{b'}=1}[\emb{p}P]_{b'}\min_{\{\emb{\alpha}_i\}}\emb{\alpha}_i(b')
+\sum_{b'}\indic{u_{b'}=0}[\emb{p}P]_{b'}\min_{\{\emb{\alpha}_i\}}\emb{\alpha}_i(b')\Bigg\}\nonumber\\
&=\min_{\emb{u}\in\{0,1\}^n}\Bigg\{\sum_{i=1}^n [\emb{p}P]_i\left(\indic{u_i=0}+\sum_{\ell=1}^n c \indic{u_{\ell}=1}\right)+\sum_{b'} [\emb{p}P]_{b'}\min \emb{\alpha}_i\cdot \emb{e}_{b'}\Bigg\}\nonumber\\
&= \min_{\emb{u}\in\{0,1\}^n}\Bigg\{\sum_{i=1}^n [\emb{p}P]_i\left(\indic{u_i=0}+\sum_{\ell=1}^n c \indic{u_{\ell}=1}\right)+\sum_{b'} [\emb{p}P]_{b'}J(\emb{e}_{b'})\Bigg\}
\end{align}
\noindent Interchanging the order of the summation and minimization corresponds to a fully observable state after the next scheduling action, i.e., that the future belief is $\emb{e}_{b'}$. Hence, the \QMDP{} value function is a lower bound on the cost function of the original problem.

Unfortunately, this is only true for the simplistic model and does not extend to the coupled models since the factored tracking cost in (\ref{eq:artif_decomp}) need not be a lower bound on the true tracking cost.

To obtain a lower bound on the optimal energy-tracking tradeoff for such models, we combine the observable-after-control assumption with a decomposable lower bound on the tracking cost which we derive next. Consider the continuous observation model with discrete state space. Given the current belief $\emb{p}_k$ and a control vector $\emb{u}_k$ the expected tracking cost can be written as:
\begin{eqnarray}
E[d(\hat{b}_{k+1},b_{k+1})|\emb{p}_k,\emb{u}_k]&=& \sum_{j=1}^m \Pr[\hat{b}_{k+1}\ne j|\emb{p}_k,\emb{u}_k,b_{k+1}=j]\Pr[b_{k+1}=j|\emb{p}_k,\emb{u}_k]\nonumber\\
&=& \sum_{i=1}^m \emb{p}_k(i)\sum_{j=1}^m p(b_{k+1}=j|b_k=i)\Pr[\hat{b}_{k+1}\ne j|\emb{p}_k,\emb{u}_k,b_{k+1}=j]\nonumber\\
\end{eqnarray}
Defining
\[
P(E|H_j)\triangleq \Pr[\hat{b}_{k+1}\ne j|\emb{p}_k,\emb{u}_k,b_{k+1}=j]
\]
\noindent which is a conditional error probability for a multiple hypothesis testing problem with $m$ hypotheses, each corresponding to a different mean vector contaminated with white Gaussian noise. Conditioned on $H_j$ the observation model is:
\begin{equation}
H_j: \emb{s}(\ell) = (\emb{m}_j(\ell) + \emb{w}(\ell))\indic{u_{k,\ell}=0} + \varepsilon \indic{u_{k,\ell}>0}
\label{eq,underHj}
\end{equation}
\noindent where $\emb{s}(\ell)$ is the $\ell$-th entry of an $n\times 1$ vector $\emb{s}$ denoting the received signal strength at the $n$ sensors, $\emb{m}_j$ is the mean received signal strength when the target is at state $j$ ($j$-th hypothesis), and $\emb{w}$ is a zero mean white Gaussian Noise, i.e. $\emb{w}\sim {\cal N}(0,\sigma^2I)$. According to  (\ref{eq,underHj}), sensor $\ell$ gets a Gaussian observation, which depends on the future target location, if activated at the next time step,  and an erasure, otherwise. Since the current belief is $\emb{p}_k$, the prior for the $j$-th hypothesis is $\pi_j=[\emb{p}_k P]_j$. The error event $E$ can be written as the union of pairwise error regions as
\begin{eqnarray}
p(E|H_j) = \Pr[\cup_{k\ne j}\zeta_{kj}]
\end{eqnarray}
\noindent where
\[
\zeta_{kj}= \{\emb{s}:L_{kj}(\emb{s})>\frac{\pi_j}{\pi_k}\}
\]
is the region of observations for which the $k$-th hypothesis $H_k$ is more likely than the $j$-th hypothesis $H_j$ and where
\[
L_{kj}\triangleq \frac{f(\emb{s}|H_k)}{f(\emb{s}|H_j)}
\]
denotes the likelihood ratio for $H_k$ and $H_j$. Using standard analysis for likelihood ratio tests \cite{poor,levy}, it is not difficult to show that:
\begin{equation}
 p(\zeta_{kj}|H_j)=Q\left(\frac{d_{kj}}{2}+\frac{\ln\frac{\pi_j}{\pi_k}}{d_{kj}}\right)
\end{equation}
\noindent where, $d^2_{kj}=\frac{\emb{\Delta m}_{kj}^T\emb{\Delta m}_{kj}}{\sigma^2}$, $\emb{\Delta m}_{kj}=\emb{m}_k-\emb{m}_j$, and $Q(.)$ is the normal distribution $Q$-function. The quantity $d_{kj}$ plays the role of distance between the two hypothesis and hence depends on the difference of their corresponding mean vectors and the noise variance $\sigma^2$. Note that, for different values of $k$ and $j$, $\zeta_{kj}$ are not generally disjoint but allow us to lower bound the error probability in terms of pairwise error probabilities, namely, a lower bound can be written as:
\begin{eqnarray}
p(E|H_j)\geq\max_{k\ne j} p(\zeta_{kj}|H_j).
\end{eqnarray}
And we can readily lower bound the expected tracking error:
\begin{eqnarray}
E[d(\hat{b}_{k+1},b_{k+1})|\emb{p}_k,u_k]&\geq&\sum_{i=1}^m \emb{p}_k(i)\sum_{j=1}^m p(b_{k+1}=j|b_k=i)\max_{k\ne j} p(\zeta_{kj}|H_j)\nonumber\\
& = &\sum_{i=1}^m \emb{p}_k(i)\sum_{j=1}^m p(b_{k+1}=j|b_k=i)\max_{k\ne j}Q\left(\frac{d_{kj}}{2}+\frac{\ln\frac{\pi_j}{\pi_k}}{d_{kj}}\right)
\label{eq:lb_tracking}
\end{eqnarray}
Next we separate out the effect of each sensor on the tracking error:
\begin{eqnarray}
E[d(\hat{b}_{k+1},b_{k+1})|\emb{p}_k,\emb{u}_k]&\geq&\indic{u_{k,\ell}=1}E[d(\hat{b}_{k+1},b_{k+1})|\emb{p}_k,\emb{u}_{k}=\emb{1}]\nonumber\\
&+& \indic{u_{k,\ell}=0}E[d(\hat{b}_{k+1},b_{k+1})|\emb{p}_k,u_{k,i}=0 ~~\forall i\ne\ell]~~\mbox{for every}~~\ell\nonumber\\
\label{eq,trick}
\end{eqnarray}
\noindent where $\emb{1}$ is the vector of all ones designating that all sensors will be active at the next time slot. The inequality in (\ref{eq,trick})  follows from the fact that if we separate out the effect of the $\ell$-th sensor we get a better tracking performance when all the remaining sensors are awake. Since this holds for every $\ell$, a lower bound on the expected tracking error can be written as a convex combination of all sensors contributions:
\begin{align}
E[d(\hat{b}_{k+1},b_{k+1})|\emb{p}_k,\emb{u}_k]\geq
\sum_{\ell=1}^n \lambda_{\ell}(\emb{p}_k)\Big\{&\indic{u_{k,\ell}=1}E[d(\hat{b}_{k+1},b_{k+1})|\emb{p}_k,\emb{u}_{k}=\emb{1}]\nonumber\\
&+ \indic{u_{k,\ell}=0}E[d(\hat{b}_{k+1},b_{k+1})|\emb{p}_k,u_{k,i}=0 ~~\forall i\ne\ell]\Big\}
\end{align}
\noindent where, $\sum_{\ell}\lambda_{\ell}(\emb{p}_k) = 1$.

Let $\emb{1}_{-\ell}$ denote a vector of length $n$ with all entries equal to one except for the $\ell$-th entry being zero. Then replacing from (\ref{eq:lb_tracking}),
\begin{align}
E[d(\hat{b}_{k+1},b_{k+1})|&\emb{p}_k,\emb{u}_k]\geq\nonumber\\
\sum_{\ell=1}^n\lambda_{\ell}(\emb{p}_k)\Bigg\{&\indic{u_{k,\ell}=1}\sum_{i=1}^m \emb{p}_k(i)\sum_{j=1}^m p(b_{k+1}=j|b_k=i)\max_{k\ne j}Q\left(\frac{d_{kj}(\emb{1})}{2}+\frac{\ln\frac{\pi_j}{\pi_k}}{d_{kj}(\emb{1})}\right)\nonumber\\
+&\indic{u_{k,\ell}>0}\sum_{i=1}^m \emb{p}_k(i)\sum_{j=1}^m p(b_{k+1}=j|b_k=i)\max_{k\ne j}Q\left(\frac{d_{kj}(\emb{1}_{-\ell})}{2}+\frac{\ln\frac{\pi_j}{\pi_k}}{d_{kj}(\emb{1}_{-\ell})}\right)\Bigg\}
\end{align}

To simplify notation, we define the following $2$ quantities:
\[
T_1(\emb{p};i,\ell)\triangleq\sum_{j=1}^m p(b_{k+1}=j|b_k=i)\max_{k\ne j}Q\left(\frac{d_{kj}(\emb{1})}{2}+\frac{\ln\frac{[\emb{p}P]_j}{[\emb{p}P]_k}}{d_{kj}(\emb{1})}\right)
\]
\[
T(\emb{p};i,\ell) \triangleq\sum_{j=1}^m p(b_{k+1}=j|b_k=i)\max_{k\ne j}Q\left(\frac{d_{kj}(\emb{1}_{-\ell})}{2}+\frac{\ln\frac{[\emb{p}P]_j}{[\emb{p}P]_k}}{d_{kj}(\emb{1}_{-\ell})}\right)
\]
\noindent Intuitively, $T_1(\emb{p};i,\ell)$ represents the contribution of sensor $\ell$ to the total expected tracking cost when the underlying state is $i$, the belief is $\emb{p}$, and when all sensors are awake. On the other hand $T(\emb{p};i,\ell)$ is the $\ell$-th sensor contribution when it is inactive and all the other sensors are awake.

Now if we assume that the target will be perfectly observable after taking the scheduling action, a lower bound on the total cost can be readily obtained from the solution of the following Bellman equation:
\begin{equation}
J(\emb{p}) = \sum_{\ell} J^{(\ell)}(\emb{p})
\label{eq,decomposed_valfunc}
\end{equation}
\noindent where
\begin{align}
J^{(\ell)}(\emb{p})=\min_{u_{\ell}\in\{0,1\}}\Bigg(&\indic{u_{\ell}=1}\left(\sum_b \emb{p}(b)\lambda_{\ell}T_1(\emb{p};b,\ell)+ c\sum_{i=1}^m [\emb{p}P]_i\right)\nonumber\\
&+\indic{u_{\ell}=0}\sum_b \emb{p}(b)\lambda_{\ell}T(\emb{p};b,\ell)+\sum_{i=1}^m [\emb{p}P]_i J^{(\ell)}(\emb{e}_i)\Bigg)
\label{eq:lb_valfunc_per_sensor}
\end{align}

Note that if we can solve the equation above for $\emb{p}=\emb{e}_i$ for all $i\in\{1,\ldots,m\}$, then it is straightforward to find the solution for all other values of $\emb{p}$. We therefore focus on specifying the value function at those points. Since this is the case, we further simplify our notation and use $T(i,\ell)$ and $\lambda(i,\ell)$ as shorthand for $T(\emb{e}_i;i,\ell)$ and $\lambda_{\ell}(\emb{e}_i)$, respectively. We can see that a lower bound on the value function of sensor $\ell$ can be obtained as a solution to the following minimization problem over $u$:
\begin{align}
J^{(\ell)}(\emb{e}_b) = \min\left\{\lambda(b,\ell)T(b,\ell);\lambda(b,\ell)T_1(b,\ell)+ c\sum_{i=1}^m [\emb{e}_bP]_i\right\}+\sum_{i=1}^m [\emb{e}_bP]_i J^{(\ell)}(\emb{e}_i)
\label{eq,final_lb_per_sensor}
\end{align}

Equation (\ref{eq,final_lb_per_sensor}) together with (\ref{eq,decomposed_valfunc}) define a lower bound on the total expected cost. To further tighten the bound we can now optimize over a matrix $\Lambda$ for every value of $c$, where $\Lambda(c)$ is an $m\times n$ matrix with the $(i,\ell)$ entry equal to $\lambda(i,\ell)$, i.e., $\Lambda(c)=\{\lambda(i,\ell)\}$. Hence
\begin{align}
J(\emb{e}_b) = \max_{\Lambda(c)}\sum_{\ell = 1}^n\left(\min\left\{\lambda(b,\ell)T(b,\ell);\lambda(b,\ell)T_1(b,\ell)+ c\sum_{i=1}^m [\emb{e}_bP]_i\right\}+\sum_{i=1}^m [\emb{e}_bP]_i J^{(\ell)}(\emb{e}_i)\right)
\end{align}
\[
\mbox{subject to}~~~~ \Lambda \emb{1}_n = \emb{1}_m
\]
\noindent where $\emb{1}_m$ is a column vector of all ones of length $m$.

\noindent The inner recursion can be solved to obtain a closed form solution for $J^{(\ell)}(\emb{e}_b)$ as:
\begin{equation}
J^{\ell}(\emb{e}_b)=\sum_{j=0}^{\infty}\sum_{i=1}^m \min\Big\{[\emb{e}_bP^j]_i\lambda(i,\ell)T_1(i,\ell)+c\sum_{k=1}^m[\emb{e}_bP^{j+1}]_k~;~[\emb{e}_bP^j]_i\lambda(i,\ell)T(i,\ell)\Big\}
\end{equation}

Since the problem is only constrained across the different sensors, we obtain a lower bound from the solution of the following optimization problem,
\begin{equation}
\sum_{i=1}^m\max_{\lambda(i,\ell)}\sum_{\ell=1}^n\sum_{j=0}^{\infty}
[\emb{e}_bP^j]_i\min\left(\lambda(i,\ell)T_1(i,\ell)+c\sum_{k=1}^m[\emb{e}_iP]_k~;~\lambda(i,\ell)T(i,\ell)\right)
\end{equation}
subject to
\[
\sum_{\ell=1}^n \lambda(i,\ell) = 1~~\forall i = 1,\ldots,m.
\]

We observe that for every $i$ we are maximizing a concave piecewise linear function in $\lambda(i,\ell)$. We pose an equivalent convex optimization problem by realizing that the minimum of a set of concave
functions is also concave. Since affine functions are concave, we can apply the technique here. Since the problem is unconstrained across the $i$ dimension we focus on solving the max-min problem for a fixed $i$. The final solution can then be obtained by summing the objective function for $m$ subproblems.

For each $\ell = 1,\ldots, n$ add a variable $t_{\ell}$ to the optimization problem. Also for every $\ell$ append $2$ constraints to the optimization problem. The constraints state the minimization over $u_{\ell}$ implicitly, by requiring that $\lambda(i,\ell)T_1(i,\ell)+c\sum_{k=1}^m[\emb{e}_iP]_k\geq t_{\ell}$ and $\lambda(i,\ell)T(i,\ell)\geq t_{\ell}$. The modified problem is therefore:
\begin{equation}
\begin{array}{ll}
\textrm{maximize}_{\lambda(i,\ell),t_{\ell};\ell=1,\dotsc,n}&\displaystyle\sum_{\ell=1}^n t_{\ell},\\
\textrm{subject to}& \displaystyle\sum_{l=1}^n \lambda(i,\ell) \leq 1,\\
& \displaystyle\lambda(i,\ell)T_1(i,\ell)+c\sum_{k=1}^m[\emb{e}_iP]_k\geq t_{\ell},\\
& \displaystyle\lambda(i,\ell)T(i,\ell)\geq t_{\ell} , ~~~~~~~~~~~~~~~~~~~~~~~~~\ell=1,\dotsc,n.
\end{array}
\end{equation}
\noindent which can be readily solved using standard convex optimization techniques \cite{boyd}.

\section{Results and Simulations}
\label{sec:results}
In this section, we show experimental results illustrating the performance of the proposed scheduling policies for the different models considered in this paper. In each simulation run, the object was initially placed at the center of the network and the simulation run concluded when the object reached the absorbing state $\tau$. We perform Monte Carlo runs to compute the average tracking and energy costs for different values of the energy parameter $c$. For the planning phase in case of point-based policies, beliefs are sampled by simulating multiple object trajectories through the sensor network. Each trajectory starts from a random state sampled from the initial belief, picking actions at random, until the target leaves the network.

\begin{figure}
\centering
\begin{tabular}{cc}
\epsfig{file=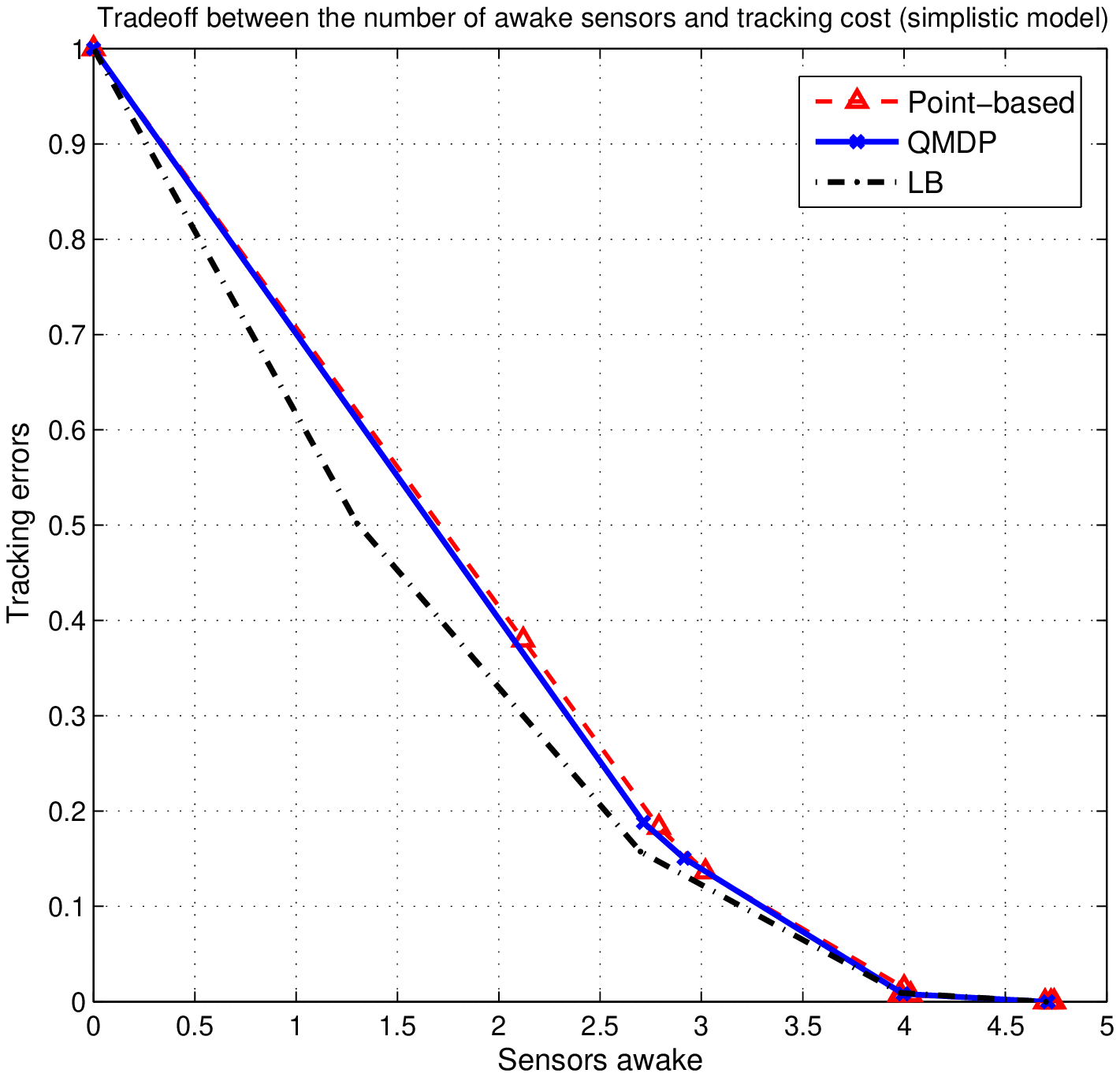,width=0.4\linewidth} &
\epsfig{file=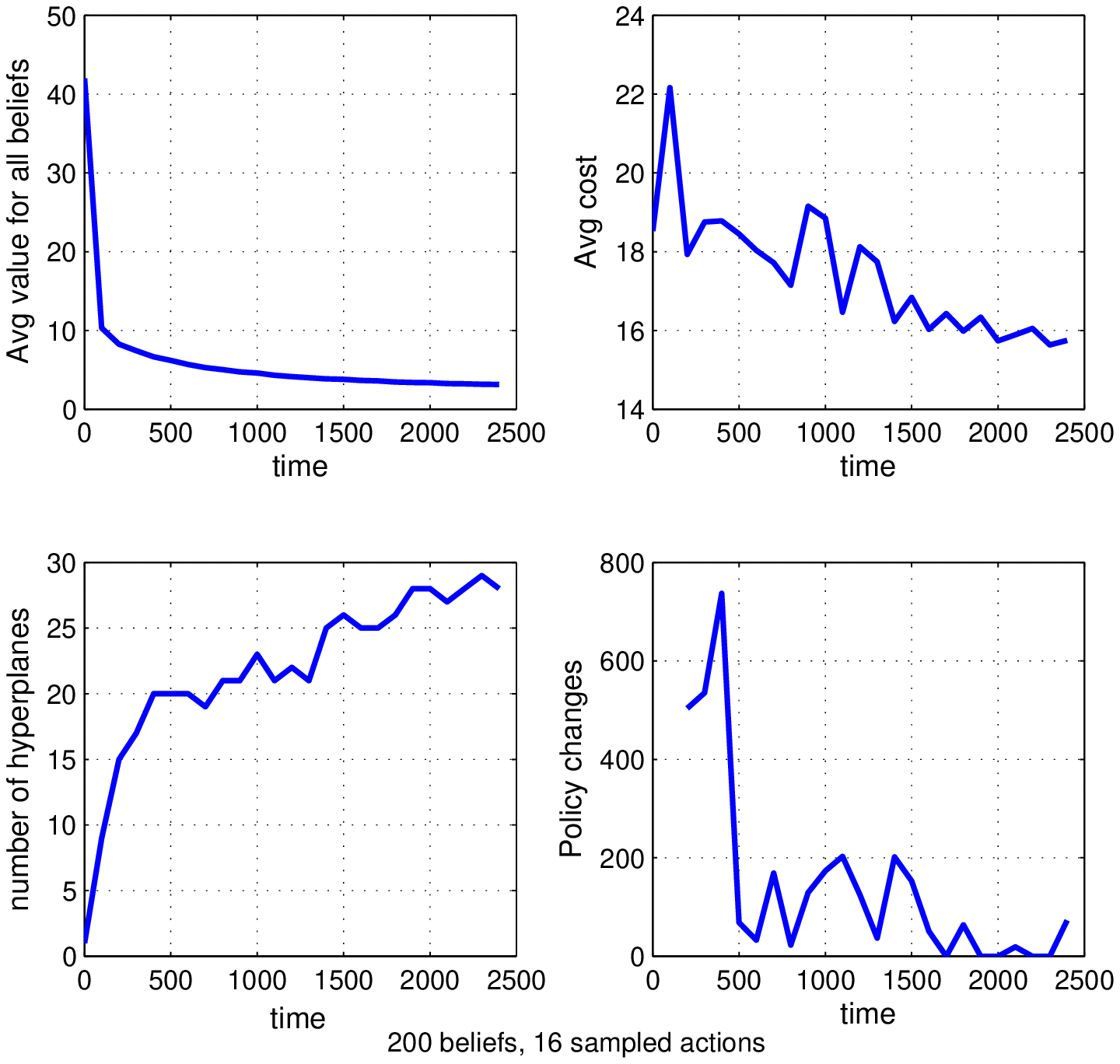,width=0.45\linewidth} \\
\mbox{(a) Energy-tracking tradeoff} & \mbox{(b) Convergence results}
\end{tabular}
\caption{Simplistic model}
\label{fig:simple_model}
\end{figure}

First, we consider the simple model in Section \ref{sec:simple_model} with a linear network of $41$ sensors. Figure \ref{fig:simple_model}(a) shows the tradeoff curve between the number of active sensors per unit time and the tracking error per unit time using the point-based and the \QMDP{} policies. The figure also shows a lower bound on the optimal performance (see Section \ref{sec:lower_bounds}). It is clear that both policies lead to tradeoffs that closely approach the lower bound. The \QMDP{} policy gets even closer to the lower bound at small tracking errors since the observable-after-control assumption is more meaningful in this regime. In Fig. \ref{fig:simple_model}(b) we show convergence results for the point-based algorithm with reduced control space minimization. The top left subplot displays the convergence of the sum cost of all the belief points in $\cal P$; the top right shows the expected cost averaged over many trajectories; the bottom left subplot shows the number of hyper-planes constituting the value function as a function of time; the bottom right subplot shows the number of policy changes versus time, i.e., the number of belief points for which the optimal action changed over $2$ consecutive iterations of the algorithm.

\begin{figure*}
\centering
\includegraphics[width=8cm]{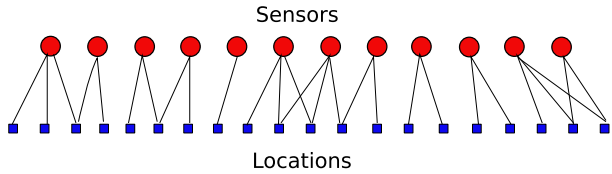}
\caption{A sensor network with overlapping sensing ranges (12 sensors and 20 object locations). An edge connects a sensor to a given location if this location falls within the sensing range of that sensor.}
\label{fig:overlap_net}
\end{figure*}

\begin{figure*}
\centering
\includegraphics[width=17cm]{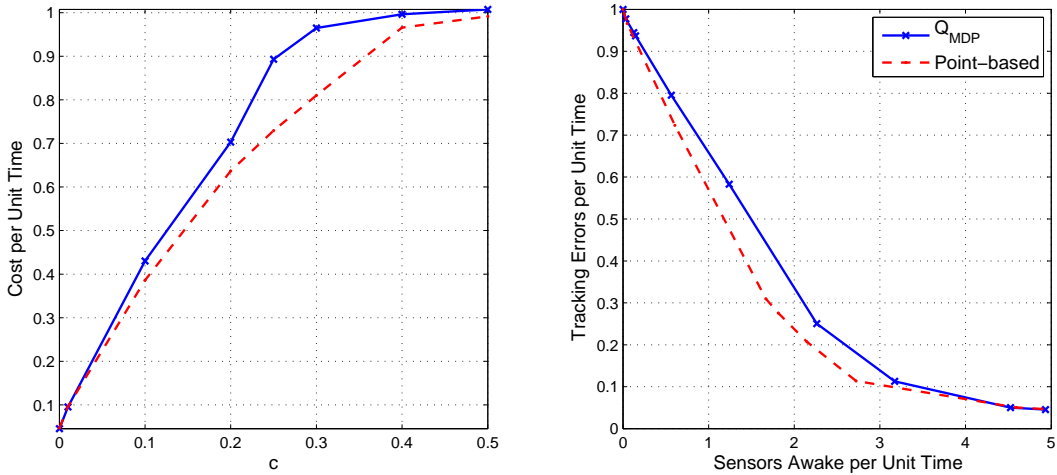}
\caption{Overlap model}
\label{fig:overlap_model}
\end{figure*}

Figure \ref{fig:overlap_model} displays the tradeoff curves for the network in Fig. \ref{fig:overlap_net} with a probabilistic observation model. The network is composed of 12 sensors and 20 object locations with the shown connectivity such that the observation range for the different sensors overlap. Since the tracking error for this model is inherently coupled across sensors, the global point-based policy clearly outperforms the learning-based \QMDP{} policy.

\begin{figure}
\centering
\begin{tabular}{cc}
\epsfig{file=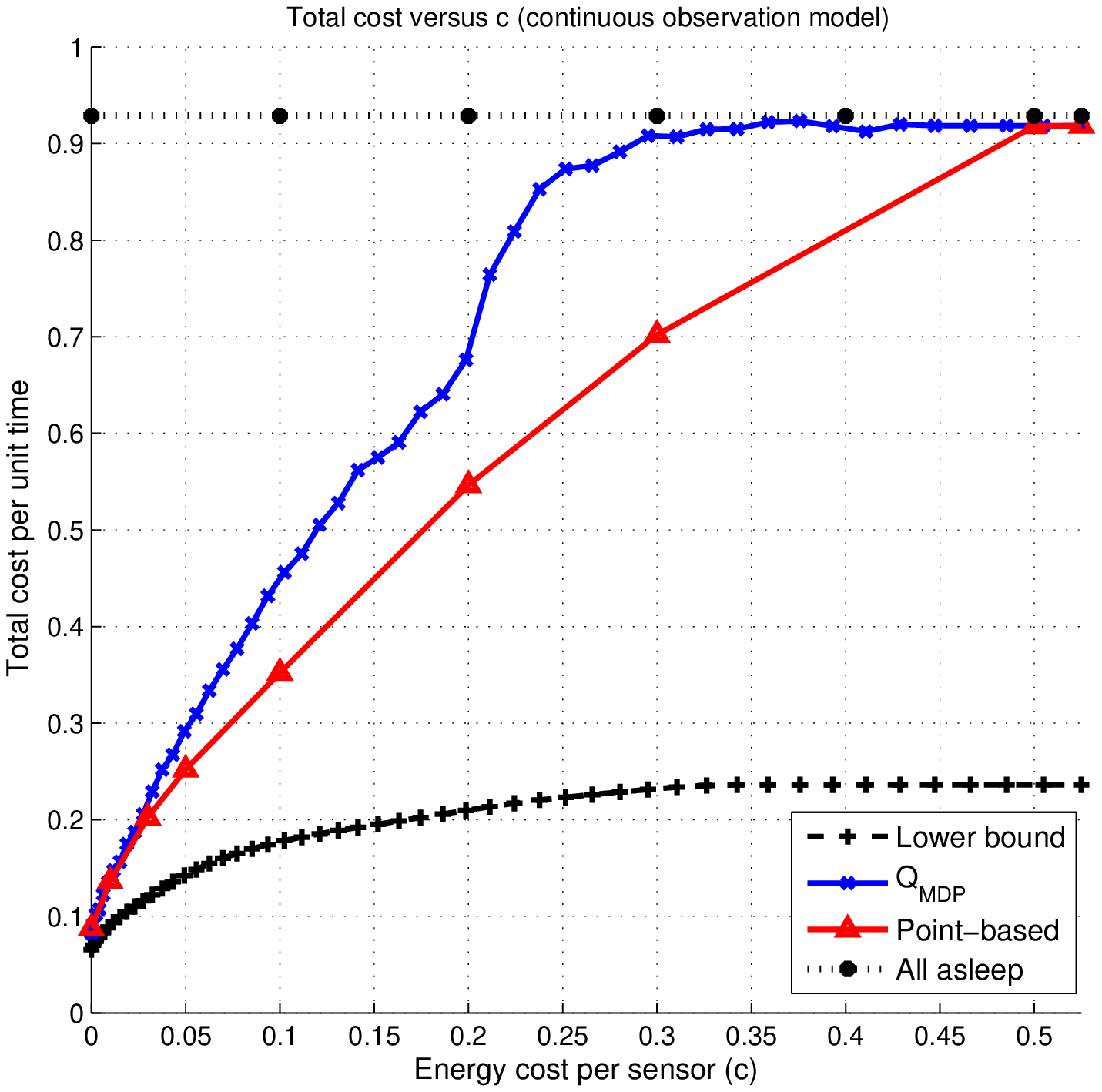,width=0.48\linewidth} &
\epsfig{file=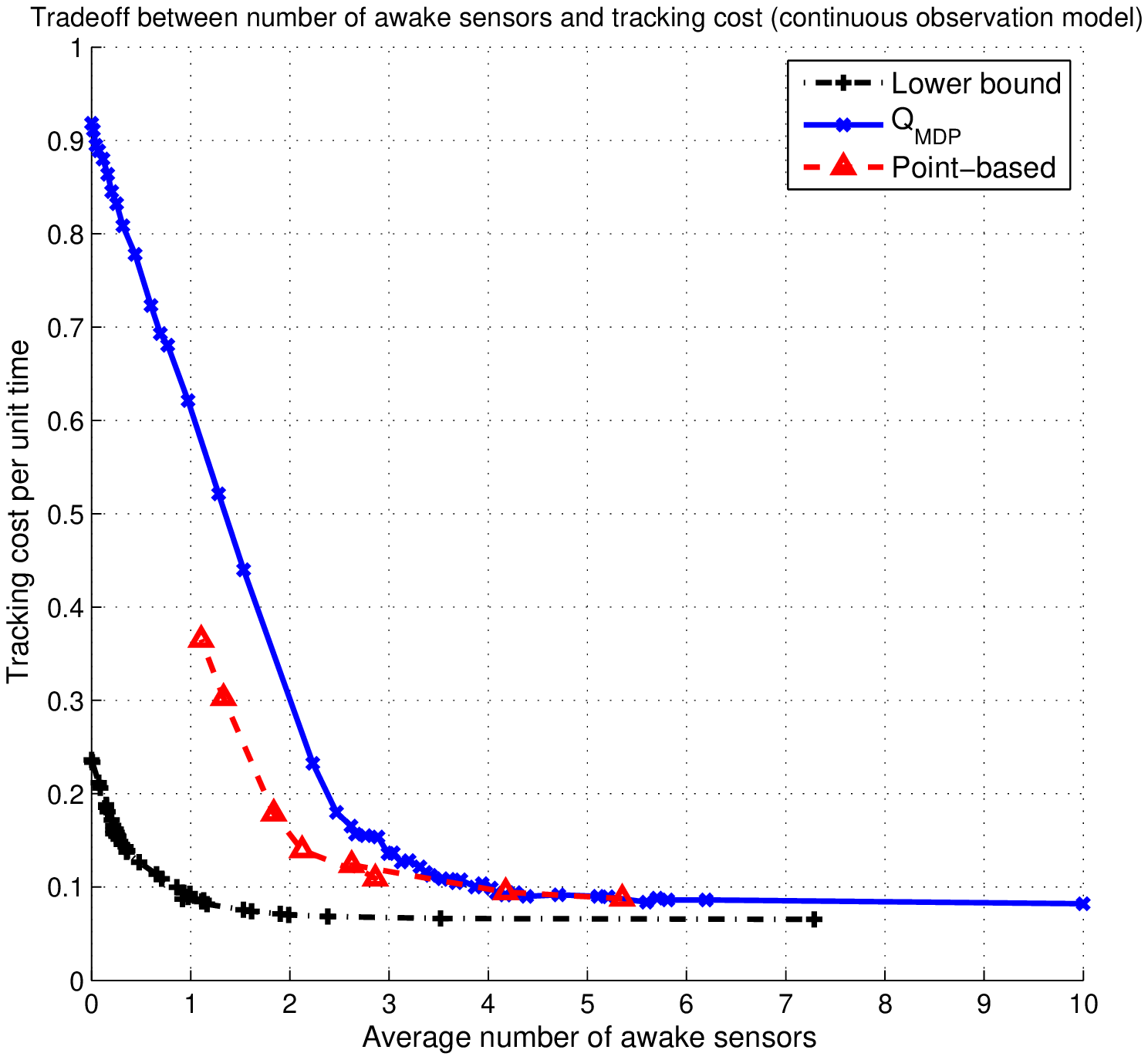,width=0.48\linewidth} \\
\mbox{(a)} & \mbox{(b)}
\end{tabular}
\caption{Continuous observation model: (a) Total cost versus energy cost per sensor, (b) Energy-tracking tradeoff}
\label{fig:DS_CO}
\end{figure}

Next, we consider a network of $10$ sensors where object locations are located on integers from $1$ to $21$. The observation for each sensor is continuous as in (\ref{cont_obs_model}). For every object state and every scheduling action in the reduced control space, we sample $50$ observations to construct estimates of the weight probabilities and compute the aggregate observation boundaries. Up to $32$ actions are sampled from the reduced control space. In this setup, the belief set consists of $500$ sampled belief vectors and we assume a Hamming error cost. Fig. \ref{fig:DS_CO} shows the performance of the different policies for the continuous observation model. It is shown that the point-based scheduling policy outperforms the \QMDP{} policy. We further show a lower bound on the optimal performance tradeoff. The lower bound is loose especially in the high tracking error regime since the derived bound on per-sensor tracking errors assumes all other sensors are awake. However, we can exactly compute the saturation point for the optimal scheduling policy since every policy has to eventually meet the all-asleep performance curve, shown in Fig.~\ref{fig:DS_CO}a,  when the energy cost per sensor is high. At that point, all sensors are inactive and hence the target estimate can only be based on prior information.

\section{Conclusions}
\label{sec:conclusions}
In this paper we studied the problem of tracking an object moving randomly through a dense network of
wireless sensors. We devised approximate strategies for scheduling the sensors to optimize the tradeoff
between tracking performance and energy consumption for a wide range of models. First, we proposed policies that rely on an observable-after-control assumption (\QMDP{} policies). Key to this solution is the decoupling of the optimization problem into per-sensor subproblems combined with simulation-based learning of individual tracking costs for each subproblem. Second, we developed point-based sensor scheduling strategies which optimize the value function over a small set of reachable beliefs within the belief simplex. Based on the belief support and the sparsity of the transition models, we developed a methodology to sample actions from reduced control spaces. This was combined with observation aggregation to address the complexity of the observation space for continuous observations models. In some cases we derived lower bounds on the optimal tradeoff curves. While being suboptimal, the generated scheduling policies often provide close-to-optimal energy-tracking tradeoffs. 
Developing distributed scheduling strategies when no central controller is available is an area for future research. Another interesting challenge is when the statistics for object movement are unknown or partially known.


\bibliographystyle{IEEEbib}
\bibliography{IEEEabrv,tsp_references}
\end{document}